\documentclass[letterpaper,11pt]{article}
\pdfoutput=1
\usepackage{amsmath,amssymb,bbm}
\usepackage{epsfig}
\usepackage[nosort]{cite}
\usepackage{multirow}

\begin{document}

\baselineskip=18pt


\thispagestyle{empty}
\vspace{20pt}
\font\cmss=cmss10 \font\cmsss=cmss10 at 7pt


\hfill
\vspace{20pt}

\begin{center}
{\Large \textbf
{
Anomalous Couplings in Double Higgs Production
}}
\end{center}

\vspace{15pt}
\begin{center}
{\large Roberto Contino$^{\, a}$, Margherita Ghezzi$^{\, a}$, Mauro Moretti$^{\, b}$, \\[0.2cm] Giuliano Panico$^{\, c}$,
Fulvio Piccinini$^{\, d}$ and Andrea Wulzer$^{\, c,e}$} 

\vspace{25pt}
\centerline{$^{a}$ {\small \it Dipartimento di Fisica, Universit\`a di Roma ``La Sapienza" and INFN, Sezione di Roma, Italy}}
\vskip 3pt
\centerline{$^{b}$ {\small \it Dipartimento di Fisica, Universit\`a di Ferrara and INFN, Sezione di Ferrara, Italy}}
\vskip 3pt
\centerline{$^{c}$ {\small \it  Institute for Theoretical Physics, ETH, CH-8093, Zurich, Switzerland}}
\vskip 3pt
\centerline{$^{d}$ {\small \it INFN, Sezione di Pavia, Italy}}
\vskip 3pt
\centerline{$^{e}$ {\small \it Dipartimento di Fisica e Astronomia and INFN, Sezione di Padova, Italy}}

\end{center}

\vspace{20pt}
\begin{center}
\textbf{Abstract}
\end{center}
\vspace{5pt} {\small
The process of gluon-initiated double Higgs production is sensitive to  non-linear interactions of the Higgs boson.
In the context of the Standard Model, studies of this process focused on the extraction of the Higgs trilinear coupling.
In a general parametrization of New Physics effects, however, an even more interesting interaction that can  be tested through this channel 
is the $t{\overline{t}}hh$ coupling.  This interaction vanishes in the Standard Model and is a genuine signature of theories in which  the Higgs 
boson emerges from a strongly-interacting sector.  In this paper we perform a model-independent estimate of the LHC potential to detect 
anomalous Higgs couplings in  gluon-fusion double Higgs production. 
We find that while the  sensitivity to the trilinear is poor, the perspectives of measuring the new $t{\overline{t}}hh$ coupling are rather promising.
}

\vfill\eject
\noindent


\section{Introduction}

Measuring the couplings of the Higgs boson at the LHC is a difficult but important task. It will give crucial information to distinguish  
among different theoretical scenarios that can lead to a Higgs-like particle, and  can thus shed light on the mechanism behind electroweak symmetry breaking (EWSB).
In the Standard Model (SM), the request of perturbativity and unitarity up to Planckian scales fixes the strength of all the interactions
of the Higgs boson in terms of its mass. 
Sizable modifications of the couplings can arise in  weakly-coupled extensions, such as supersymmetry,
through the mixing of the Higgs boson with new light states. In this case one expects to produce these new particles directly at the collider.
A second compelling possibility is that the EWSB
is triggered by new strong dynamics at the TeV scale,  and a light
Higgs  emerges as the pseudo Nambu-Goldstone boson (pNGB) of a larger spontaneously broken symmetry~\cite{compositeHiggs}.
In this case the modification of couplings is not necessarily accompanied by the presence of new light scalars and the direct manifestation 
of New Physics can be  postponed to TeV energies.
In addition to a modified pattern of linear couplings, this scenario predicts new non-linear interactions of the Higgs to the SM fields, which 
can lead to striking signatures at the collider and are a genuine
feature of the underling strong dynamics.

The problem of extracting the Higgs couplings by measuring its  production and decay rates at the LHC has been studied at length in the literature,
see for example Refs.~\cite{Zeppenfeld:2000td,Conway:2002kk,Belyaev:2002ua,duhrssen,Duhrssen:2004cv,Lafaye:2009vr,Bock:2010nz,Bonnet:2011yx}.
The importance of a model-independent approach has been recently re-discussed, 
and a first estimate of the impact of the current LHC data on the Higgs parameter space has been performed 
in~\cite{Carmi:2012yp,Azatov:2012bz,Espinosa:2012ir,Giardino:2012ww,Rauch:2012wa,Ellis:2012rx,Azatov:2012rd,Farina:2012ea,Degrande:2012gr,Klute:2012pu}.

Aim of this paper is to study the effect of anomalous couplings in the process of gluon-initiated double Higgs production at the LHC, $gg\rightarrow hh$. Under the 
reasonable assumption of weak couplings to light fermions, this process proceeds through top quark loops and  it is thus sensitive, in the first place, to 
the Higgs-top couplings.
It also receives a contribution from the Higgs trilinear coupling, and for this reason has been studied in detail in the context of the 
SM~\cite{Glover:1987nx,Plehn:1996wb,Dawson:1998py,Baur:2002rb,Baur:2002qd,Baur:2003gpa,Baur:2003gp}.
In theories of New Physics, however, a much more interesting coupling that can be probed through this process is the non-linear interaction $t\bar t hh$.
The latter is generally present in theories of composite Higgs, like for example the minimal models
MCHM4 \cite{Agashe:2004rs} and MCHM5~\cite{Contino:2006qr}, and gives a genuine signal of the Higgs strong interactions.
As first noticed by the authors of Ref.~\cite{Grober:2010yv} 
(see also Ref.~\cite{Dib:2005re} for a discussion of the role of the $t\bar thh$ coupling in context of Little Higgs theories),
the presence of the new coupling can lead to a  dramatic increase of the  cross section.
For example, enhancements  larger than one order of magnitude are possible in the MCHM5, and even for $(v/f)^2 \sim 0.15$, where $f$ is the decay constant of the
pNGB Higgs, the total cross section \emph{doubles} compared to its SM value.
Given that the deviations due to Higgs compositeness are usually much milder,  $gg\rightarrow hh$ seems an extremely favored channel which is worth investigating. 

In this paper we derive a first quantitative assessment on the detectability of the anomalous coupling $t\bar t hh$ in the process $gg\rightarrow hh$.
We do not consider the rarer process of double-Higgs production via vector boson fusion, which has been investigated 
at the LHC in previous  studies~\cite{Moretti:2004wa,Giudice:2007fh,Contino:2010mh,Contino:2011np},
since it is sensitive to the couplings of the Higgs to vector bosons and to the Higgs trilinear coupling.  Neglecting such process is a very good approximation, 
considering that in absence of dedicated  kinematic cuts its rate at the LHC is much smaller than the rate of $gg\rightarrow hh$.
In section~2 we briefly summarize the parametrization of the couplings of a generic Higgs-like scalar which we adopt.
In section~3 we analyze the $gg\rightarrow hh$ process in the presence of modified Higgs interactions. 
In particular we study the dependence of the 
cross section on the various couplings and show that there is high sensitivity to the new $t\bar t hh$ coupling, much larger than that on the trilinear self-interaction.
For our analysis we wrote a dedicated computer code which computes the exact 1-loop matrix element for single and double Higgs production via gluon fusion as a function 
of the relevant couplings. The code has been implemented as one of the available processes of the event generator ALPGEN~\cite{alpgen} and will be made public with its
next official release. We then discuss two of the most promising decay channels of the Higgs pair: $hh \to WW\gamma\gamma \to l\nu jj \gamma\gamma$
in the case in which the Higgs has a suppressed single coupling to the top (fermiophobic limit), and $hh \to b\bar b \gamma\gamma$ in the case in which the linear 
couplings  are SM like. For the latter case, we follow the strategy proposed in Ref.~\cite{Baur:2003gp} and in section~4 we perform a first collider study to estimate the 
exclusion and discovery limits on the anomalous $t\bar t hh$ coupling. To compute the SM background cross section we use the results of~\cite{Baur:2003gp} with 
updated $b$ and  $\gamma$  efficiencies and rejection factors. We collect the results of our collider study in section~4.1. Finally, conclusions are reported in section~5.

\section{General parametrization of the Higgs couplings}

In this section we introduce the general parametrization of  Higgs couplings
that will be used in this paper.

The most general effective Lagrangian that parametrizes the  interactions of a Higgs-like scalar
at low energy has been discussed in~\cite{Contino:2010mh} and extended in~\cite{Azatov:2012bz}. 
Under the assumption
of custodial symmetry, the Nambu-Goldstone bosons associated to the electroweak
symmetry breaking can be described as the
coordinates of the coset $SU(2)_L \times SU(2)_R/SU(2)_V \sim SO(4)/SO(3)$.
They can be conveniently parametrized by the $2\times 2$ matrix
\begin{equation}
\Sigma = \exp\left(i \sigma_a \chi^a(x)/v\right)\,, \qquad \quad a=1,2,3
\end{equation}
where $\sigma^a$ are the Pauli matrices and $v = 246\ {\rm GeV}$.
At energy scales much below possible new physics states, the  effective
Lagrangian describing a light Higgs $h$ has the form
\begin{equation}
\label{eq:eff_lagr}
\begin{split}
{\cal L}  
=&\, \frac{v^2}{4} {\rm Tr}(D_\mu \Sigma^\dagger D^\mu \Sigma)
\left(1 + 2 a \, \frac{h}{v} + \ldots\right) + \frac{1}{2} (\partial_\mu h)^2 - \frac{1}{2} m_h^2 h^2
- d_3 \,\frac{1}{6} \left(\frac{3 m_h^2}{v}\right) h^3 + \ldots \\[0.2cm]
& - m_t \, \bar q_L^i \Sigma_{i1} t_R 
\left(1 + c_t \,\frac{h}{v} + c_2\,\frac{h^2}{v^2} + \ldots\right) 
- m_b \, \bar q_L^i \Sigma_{i2} b_R \left(1 + c_b \,\frac{h}{v} + \dots \right) + h.c.\, ,
\end{split}
\end{equation}
where $q_L = (t_L, b_L)$ and $a$, $c_{t,b}$, $c_2$ and $d_3$ are the numerical coefficients that parametrize the Higgs couplings.
The dots stand for terms which are not relevant for 
double Higgs production via gluon-fusion. In particular, we assume that the strength of single interactions of the Higgs to the fermions
is not extremely enhanced compared to its SM value, so that the contribution of the light fermions and the bottom quark to double Higgs production 
can be safely neglected.
The coupling to the bottom, $c_b$, is relevant only in the Higgs decay, and we will set 
\begin{equation}
c_t = c_b =c
\end{equation}
for simplicity in the following. 
We also neglect $ggh$ and $gghh$ local interactions which can be generated by new heavy states at 1-loop level, as for example scalar or fermionic 
partners of the top quark (see Ref.~\cite{Pierce:2006dh} for a study of the effect of such local interactions).

In this analysis we will freely vary the parameters that appear in the effective Lagrangian. In specific models, however, 
they can be related to each other. For example, the SM Lagrangian is obtained for 
\begin{equation}
a = c = d_3 = 1\,, \qquad \quad c_2 = 0\, .
\end{equation}
A class of theories that we will consider in the following are the  composite Higgs models based on
the symmetry pattern $SO(5)/SO(4)$ \cite{Agashe:2004rs,Contino:2006qr}.
In these models the parameter $a$ is given by
\begin{equation}
\label{eq:par_a}
a = \sqrt{1 - \xi}\,,
\end{equation}
where $\xi = v^2/f^2$ and $f$ is the Nambu-Goldstone decay constant.
The values of $c$, $c_2$ and $d_3$  depend on which $SO(5)$ representation the fermions are embedded in. 
For the two minimal choices of fermions in the spinorial (MCHM4 \cite{Agashe:2004rs}) and
fundamental (MCHM5 \cite{Contino:2006qr}) representations one gets
\begin{eqnarray}
&&c = d_3 = \sqrt{1-\xi}\,, \qquad
c_2 = -\frac{\xi}{2}\,, \hspace{3em} {\rm MCHM4,\ \ spinorial\ representation}\, ,\label{eq:par_MCHM4}\\[0.2cm]
&&c = d_3 = \frac{1-2\xi}{\sqrt{1-\xi}}\,, \qquad
c_2 = -2 \xi\,, \hspace{2.7em} {\rm MCHM5,\ \ fundamental\ representation}\, .\label{eq:par_MCHM5}
\end{eqnarray}
Equations (\ref{eq:par_a}), (\ref{eq:par_MCHM4}) and (\ref{eq:par_MCHM5}) account for the value of the Higgs couplings as due to the non-linearities of the chiral 
Lagrangian. The exchange of new heavy particles can however give further corrections to these expressions. In the following we will neglect these effects since they are 
parametrically subleading~\cite{Low:2009di}, although they can be numerically important when the top or bottom degree of compositeness becomes 
large~\cite{Azatov:2011qy}. 
This is especially justified  considering that in minimal composite Higgs models with partial compositeness these additional corrections to the couplings do not affect 
the $gg\to h$ rate because they are exactly canceled by the contribution from loops of heavy fermions, as first observed in Refs.~\cite{falkowski,Low:2009di} and explained 
in  Ref.~\cite{Azatov:2011qy}. For double Higgs production we expect this cancellation to occur only in the limit of vanishing momentum of the Higgs external 
lines. In general, numerically important contributions might come from light top partners (light custodians).
In  models with partial compositeness, where the dominant contribution to the Higgs potential comes from top loops, 
the presence of light fermionic resonances is
essential to obtain a light Higgs~\cite{Contino:2006qr,lightpartners}. In particular, 
$m_h \simeq 120-130\ {\rm GeV}$ requires top partners around or below $1\ {\rm TeV}$.
It would be interesting to analyze in detail their effects on double Higgs production.

\section{Double Higgs production via gluon fusion}
\label{sec:ggHH}

\begin{figure}
\centering
\includegraphics[width=147pt]{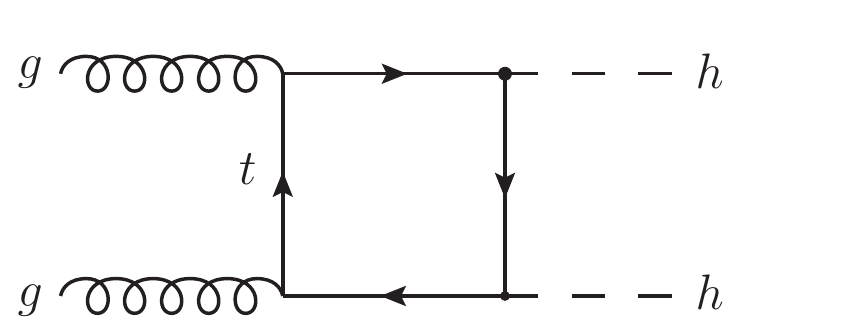}
\hspace{.5em}
\includegraphics[width=164pt]{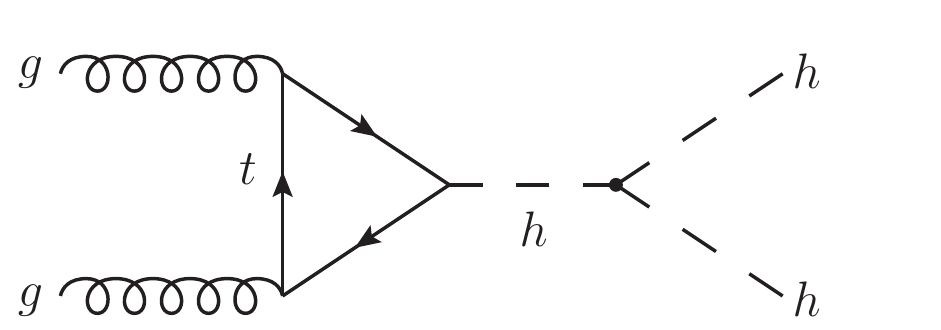}
\hspace{.5em}
\includegraphics[width=134pt]{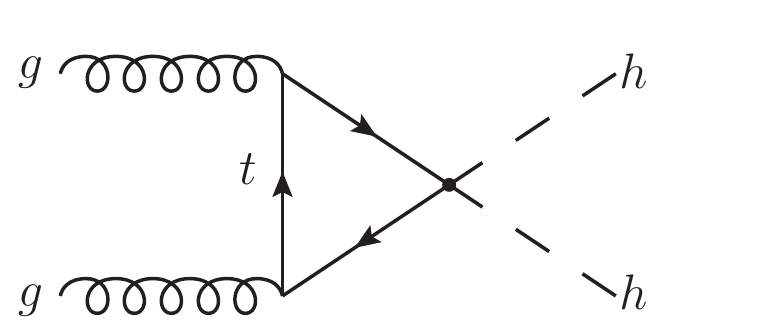}
\hspace{-2.em}
\caption{\small 
Feynman diagrams for double Higgs production via gluon fusion (an additional contribution comes from the crossing of the box diagram). 
The last diagram contains the new non-linear Higgs interaction $t\bar t hh$.}
\label{fig:diagrams}
\end{figure}

In the scenario we are considering, the leading-order contributions to the  process $gg \rightarrow hh$ come from Feynman diagrams containing
a top-quark loop. The three relevant diagrams are shown in Fig.~\ref{fig:diagrams}, and can be computed by using the results of Ref.~\cite{Plehn:1996wb}.
%
%
We have implemented the automatic computation of the matrix element as one of the processes of the ALPGEN MonteCarlo generator~\cite{alpgen}.
The code will be made public with the next official release of ALPGEN, and it allows one to compute the total cross section and differential distributions,
as well as to generate events for an arbitrary choice of the Higgs couplings $c$, $d_3$, $c_2$. The validation of the code has been performed by means of an
independent C++ program linked to the QCDLoop~\cite{Ellis:2007qk} and to the LHAPDF routines~\cite{Whalley:2005nh}.
All the results reported in the following have been derived by use of the ALPGEN
matrix element calculation with CTEQ6l parton distribution functions and renormalization and factorization scales $Q = m(hh)$. 
The top quark mass has been set to $m_t = 173\ {\rm GeV}$.

\begin{figure}[tp]
\centering
\includegraphics[width=0.5\textwidth]{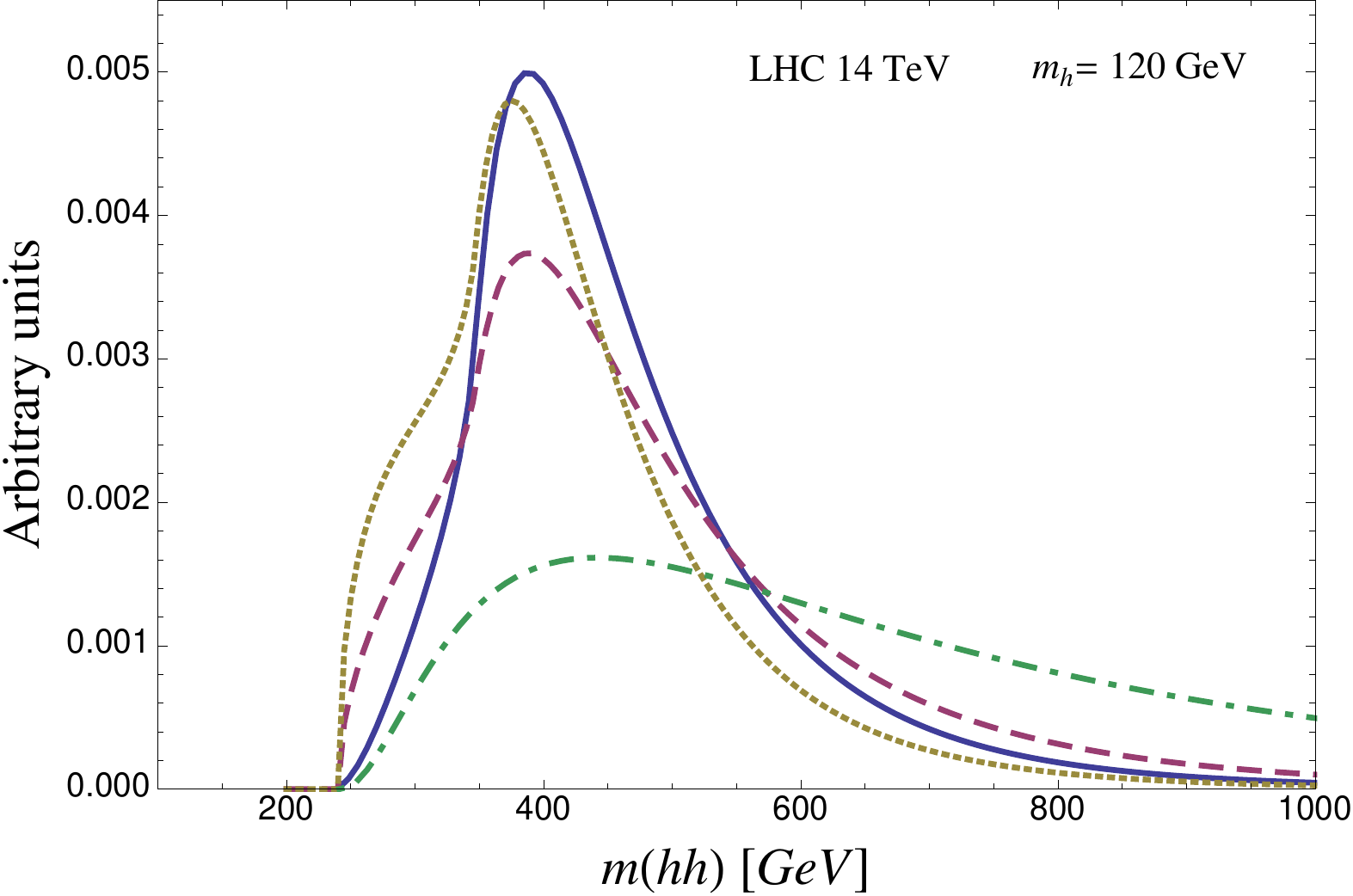}
\caption{\small
Invariant mass distribution of the two Higgs bosons in $pp \rightarrow hh$ at the LHC ($14\ {\rm TeV}$) for $m_h = 120\ {\rm GeV}$.
The various curves correspond to different choices of Higgs couplings: $c = d_3 = 1$, $c_2 = 0$ (SM couplings, solid blue curve),
$c = d_3 = 1$, $c_2 = -1$ (dashed purple curve), $c = 1$, $d_3 = c_2 = 0$ (dotted yellow curve). The dot-dashed green curve
shows the distribution obtained in the approximation of infinite top mass with SM couplings. All curves have been normalized to unit area.
The corresponding LO total cross sections are $15.2\,$fb (solid blue curve), $253\,$fb (dashed purple curve), $31.6\,$fb (dotted yellow curve).
}
\label{fig:distributions}
\end{figure}

The amplitude of each diagram in Fig.~\ref{fig:diagrams} is characterized by a different 
energy scaling at large invariant masses $\sqrt{\hat s}= m(hh) \gg m_t, m_h$. One has
\begin{align}
\label{eq:box_behaviour}
{\cal A}_\square & \sim  c^2\, \alpha_s \frac{m_t^2}{v^2}\, , \\[0.3cm]
\label{eq:triangle_d3_behaviour}
{\cal A}_\triangle & \sim  c\, d_3\, \alpha_s \frac{m_t^2}{v^2} \, \frac{m_h^2}{\hat s}
\left[\log\left(\frac{m_t^2}{\hat s}\right) + i \pi\right]^2\, , \\[0.3cm]
\label{eq:triangle_c2_behaviour}
{\cal A}_{\triangle nl} & \sim  c_2\, \alpha_s \frac{m_t^2}{v^2}
\left[\log\left(\frac{m_t^2}{\hat s}\right) + i \pi\right]^2\, ,
\end{align}
where ${\cal A}_\square$, ${\cal A}_\triangle$ are the amplitudes of respectively the box and the triangle diagram with the Higgs exchange (first two diagrams
of Fig.~\ref{fig:diagrams}), while ${\cal A}_{\triangle nl}$ denotes the amplitude of the diagram with the new non-linear interaction $t\bar t hh$ (last diagram
of Fig.~\ref{fig:diagrams}).
At large $\hat s$ the box  and the diagram with
the new vertex dominate, while the  triangle  with Higgs exchange gives its largest contribution
near threshold.
For SM values of the couplings there is a destructive interference between ${\cal A}_\triangle$ and ${\cal A}_\square$, so that decreasing  the trilinear
coupling $d_3$ leads to a softer distribution, while increasing it makes the suppression of the cross section near threshold even stronger. 
On the other hand, since ${\cal A}_{\triangle nl}$  and ${\cal A}_\square$ have similar energy scalings (the log enhancement of ${\cal A}_{\triangle nl}$ becomes
important only at very large $\sqrt{\hat s}$ where the gluon pdfs are small), 
the main effect of their interference is on the total cross section, with little modification of the  $m(hh)$ distribution.
These behaviours are clearly visible in the distributions shown in Fig.~\ref{fig:distributions} for $m_h =120\,$GeV at 14 TeV.
%
%

The cross section depends on the couplings as a quadratic polynomial in the variables $c_2$, $c^2$ and $c d_3$, associated respectively to the three diagrams 
of Fig.~\ref{fig:diagrams}. It can thus be conveniently expressed by the formula
\begin{equation}
\label{eq:fitxsectot} 
\sigma(pp\to hh) =\overline{\sigma}\left[ c_2^2+\left(\alpha\, c^2\right)^2+\left(\beta\, cd_3\right)^2+A_1\, c_2\left(\alpha\, c^2\right)
+A_2 \left(\alpha\, c^2\right)\left(\beta\, cd_3\right)+A_3\, c_2\left(\beta\, cd_3\right)\right]\,,
\end{equation}
where the value of the (real) coefficients $\overline{\sigma},\alpha,\beta,A_1,A_2,A_3$ has been
extracted  by fitting the results of a Montecarlo integration, and is reported in Table~\ref{tab:fit}  (at LO in $\alpha_s$)
for $m_h=120, 125\,$GeV at 8 and 14 TeV. 
%
\begin{table}[tb]
\centering
\small
\begin{tabular}{llcccccc}
 & & $\overline{\sigma}$ & $\alpha$ & $\beta$ & $A_1$ & $A_2$ & $A_3$ \\[0.1cm]
\hline 
\\[-0.2cm]
\multirow{2}{65pt}{$m_h = 120\,$GeV} 
     & 14 TeV \hspace{0.2cm}       & $151.3\,{\rm{fb}}$ & $0.453$  & $0.164$ & $-1.86$ & $-1.77$ & $1.66$ \\[0.15cm]
     & \hspace{0.07cm} 8 TeV       & $32.6\,{\rm{fb}}$   & $0.474$  & $0.178$ & $-1.89$ & $-1.78$ & $1.68$ \\[0.35cm]
\multirow{2}{65pt}{$m_h = 125\,$GeV} 
     & 14 TeV \hspace{0.2cm}       & $144.6\,{\rm{fb}}$  & $0.457$  & $0.169$ & $-1.85$ & $-1.79$ & $1.68$ \\[0.15cm]
     & \hspace{0.07cm} 8 TeV       & $30.5\,{\rm{fb}}$    & $0.475$  & $0.185$ & $-1.89$ & $-1.79$ & $1.70$ \\[0.35cm]
\end{tabular}
\caption{\small
Coefficients for the fit of eq.(\ref{eq:fitxsectot}) of the total LO $pp\to hh$ cross section via gluon fusion at the~LHC.
\label{tab:fit} }
\end{table}
%
In the above parametrization, the coefficients $\alpha$ and 
$\beta$ measure the sensitivity of the cross section on the parameters $c^2$ and $(c d_3)$, relative to $c_2$. 
One can see from Table~\ref{tab:fit} that the dependence on $c^2$ is significant while the one on $(c d_3)$ is rather mild.
This can be tracked back to the additional factor $(m_h^2/\hat s)$ in the amplitude of the
triangle diagram which carries the dependence on $(c d_3)$,  see eq.(\ref{eq:triangle_d3_behaviour}). 
This factor leads to a suppression at large $\hat s$ and thus, because of the kinematic threshold $\hat s>4m_h^2$, 
to a reduction of the sensitivity on $(c d_3)$ of the total cross section.

\begin{figure}[tp]
\centering
\includegraphics[width=0.49\textwidth]{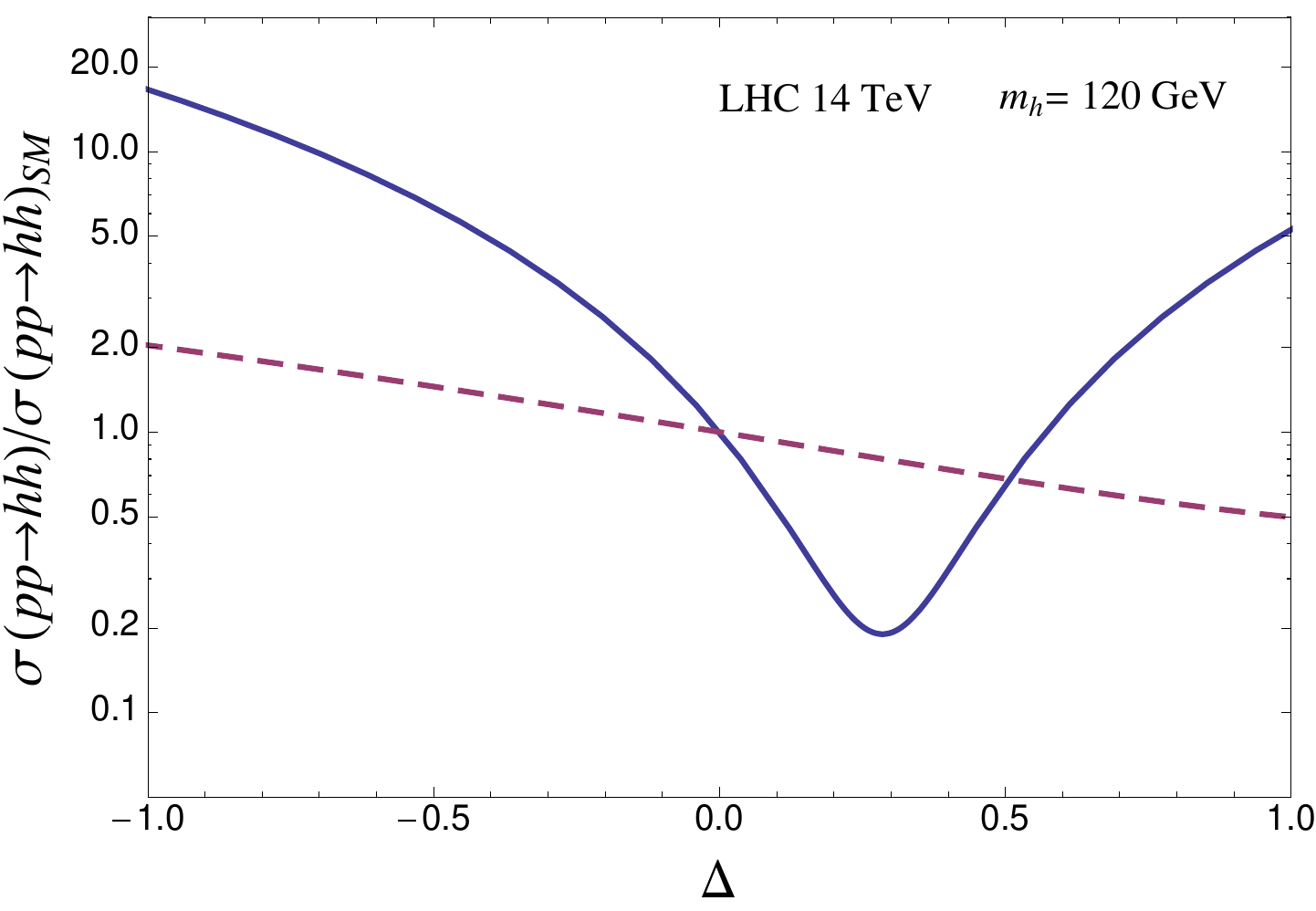} 
\includegraphics[width=0.48\textwidth]{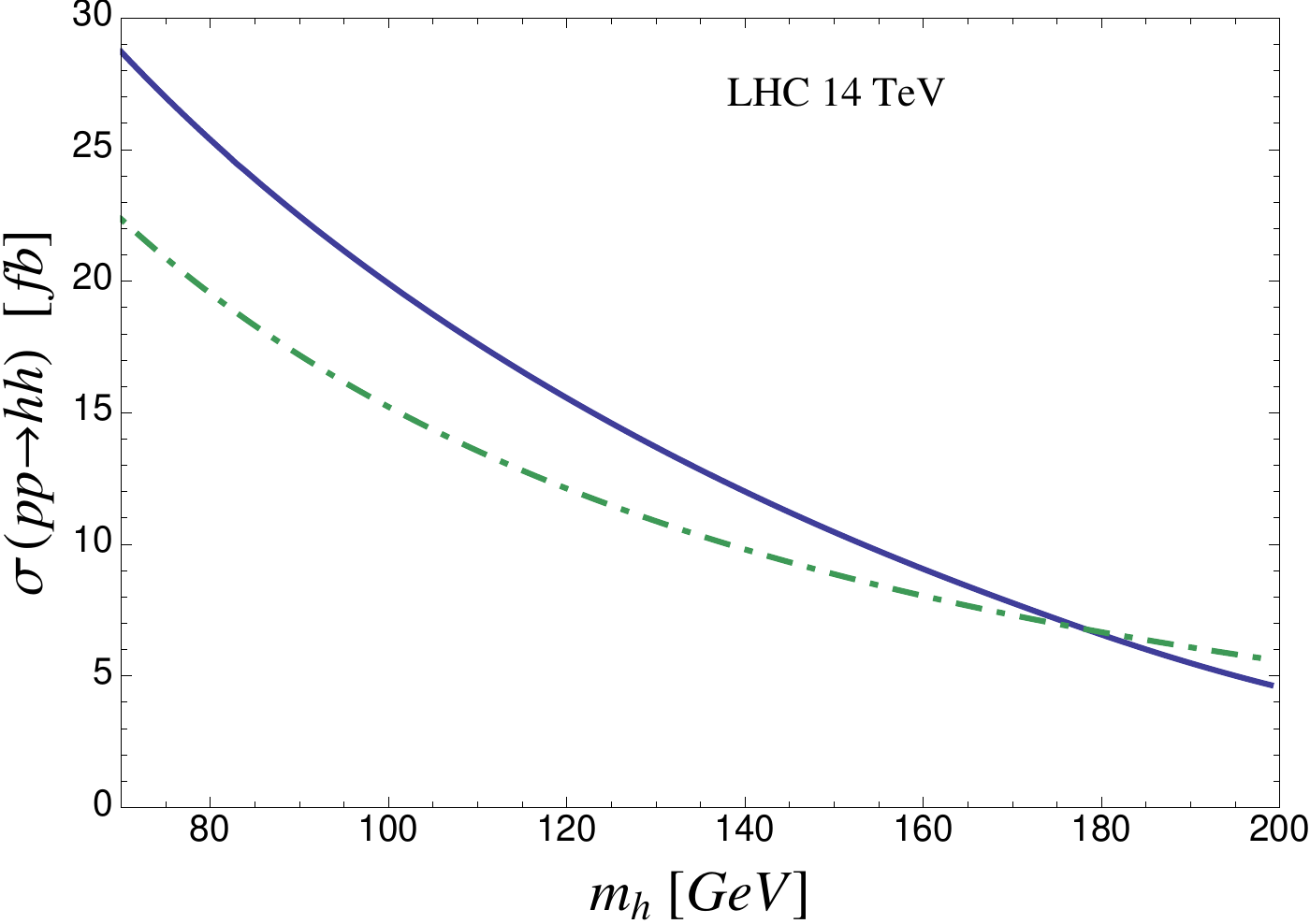}
\caption{\small
Left plot: total cross section  in SM units as a function of $c_2 = \Delta$ (solid blue curve) and $d_3 = 1 + \Delta$ (dashed purple curve), with the
other Higgs couplings set to their SM values and $m_h =120\,$GeV.
Right plot: LO total cross section (in fb) in the SM as a function of $m_h$ as computed by means of the full one-loop matrix element (solid blue curve) and the infinite
top mass approximation (dot-dashed green curve).
In both plots the LHC center-of-mass energy has been set to 14 TeV.
}
\label{fig:totalxsec}
\end{figure}

We see in Table~\ref{tab:fit} that increasing the LHC center-of-mass energy from 8 TeV to 14 TeV 
increases the LO total cross section (hence the coefficient $\bar\sigma$) by a factor $\sim 5.2$, while the relative strength among $\alpha,\beta,A_1,A_2,A_3$
varies by less than $\sim 15\%$.
When the Higgs mass is varied from $120\,$GeV to $125\,$GeV, $\bar\sigma$ decreases by $\sim 5-7 \%$, while the other coefficients change by less than $\sim 1-3\%$.
The left plot of Fig.~\ref{fig:totalxsec} illustrates how the total cross section changes when  varying individually $d_3$ and $c_2$, while fixing the other
couplings to their SM value.
%
%
In the vicinity of the SM point, decreasing (increasing) $d_3$ or $c_2$ 
leads to an enhancement (reduction) of the total cross section,
with a much stronger dependence on $c_2$ than on $d_3$.  
The right plot in the same Figure shows  how the cross section varies with the Higgs mass for the SM choice of couplings. The solid curve corresponds
to the full one-loop matrix element calculation, while the dot-dashed curve is obtained by taking the limit of infinite top mass.
As previously noticed~\cite{Baur:2002rb},~\footnote{The infinite top mass limit is also discussed by the authors of Ref.~\cite{Glover:1987nx}, although the contribution 
of the triangle diagram seems to have been accidentally omitted in their Fig.~2.} this approximation is reasonably accurate in the case of the total cross section, but 
completely fails to reproduce the correct $m(hh)$ distribution, as illustrated by the corresponding curve in Fig.~\ref{fig:distributions}.

We have seen that
modified Higgs couplings, in particular 
a non-vanishing $t\bar t hh$ interaction,  
can lead to a strong enhancement of the  total $pp\to hh$ cross section.
However, in order to determine the signal yield at the LHC in a given final state one has to take
into account also the change in the Higgs decay branching ratios.
In our case, these latter depend only on the ratio of the parameters $c$ and $a$.
We will consider two illustrative situations: \textit{i)}  the case in which the branching ratios are similar to the SM ones; 
\textit{ii)} the case in which the couplings of one Higgs boson to two fermions are suppressed (fermiophobic limit).
The first situation is realized in models where all single Higgs couplings are rescaled by the same factor (as in the MCHM4,
see eqs.(\ref{eq:par_a}), (\ref{eq:par_MCHM4})),
or  their shift from the SM value is small (as in composite Higgs models with small $\xi$).
In this case the studies of Refs.~\cite{Baur:2002rb,Baur:2002qd,Baur:2003gpa,Baur:2003gp} suggest that the most favorable final state at the LHC for a light Higgs boson 
is $hh\to \gamma\gamma b\bar b$.
On the other hand, if  single Higgs couplings to fermions are suppressed (while the strength of $t\bar thh$ can still be sizable), the dominant
decay mode for a light Higgs becomes $h\to WW$. 
The channel $hh\to WWWW$ has in this case the largest rate and should be visible in final states with two or three leptons.
The $\gamma\gamma$ branching ratio is also strongly enhanced, so that in this case  $hh\to WW\gamma\gamma$ also seems
a promising final state.
In particular, the request of one lepton from the decay of the $W$ pair should be sufficient to reduce the  background and
lead to a clean signature at the LHC.

\begin{figure}[t]
\centering
\includegraphics[width=0.45\textwidth]{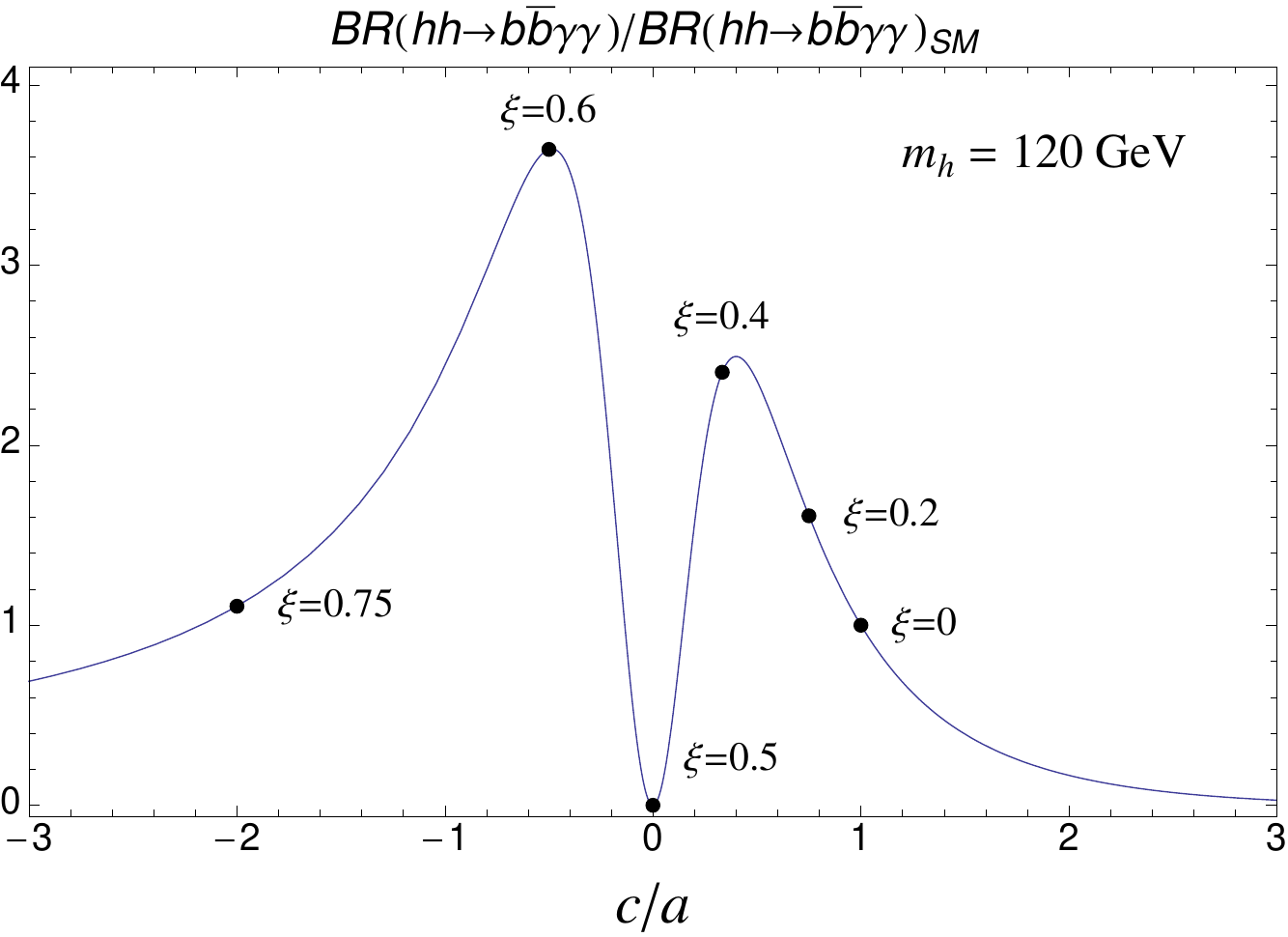} 
\includegraphics[width=0.45\textwidth]{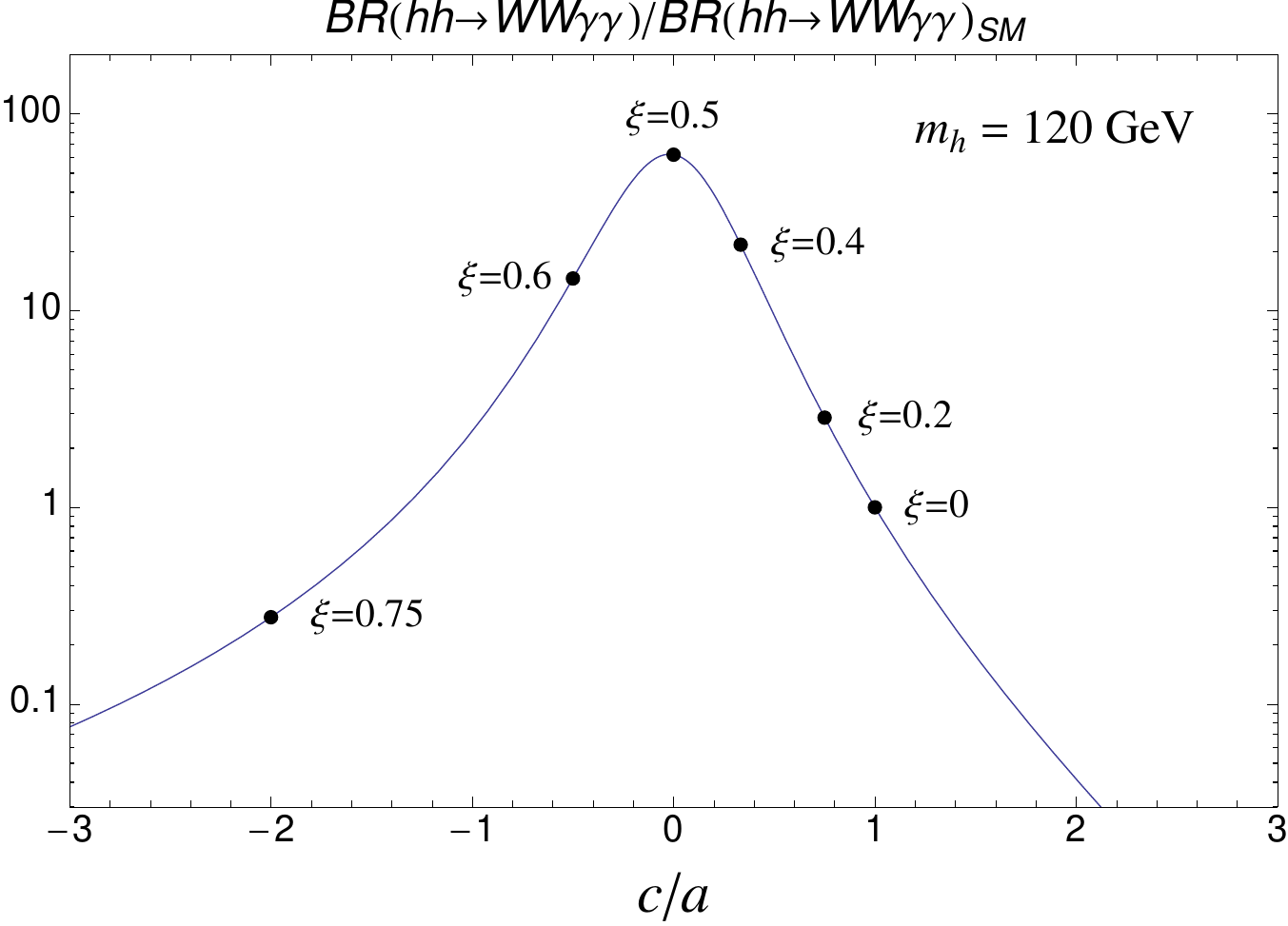} 
\caption{\small
Value of the  branching ratio $BR(hh\to \gamma\gamma b\bar b)$ (on the left) and $BR(hh\to WW\gamma\gamma)$ (on the right) in SM units 
as a function of the ratio of Higgs couplings $c/a$. The dots show the prediction in the MCHM5, where $c/a = (1-2\xi)/(1-\xi)$, for various 
values of $\xi$. In both plots the Higgs mass is set to $m_h =120\,$GeV.
}
\label{fig:BRs}
\end{figure}

\begin{figure}
\centering
\includegraphics[width=0.5\textwidth]{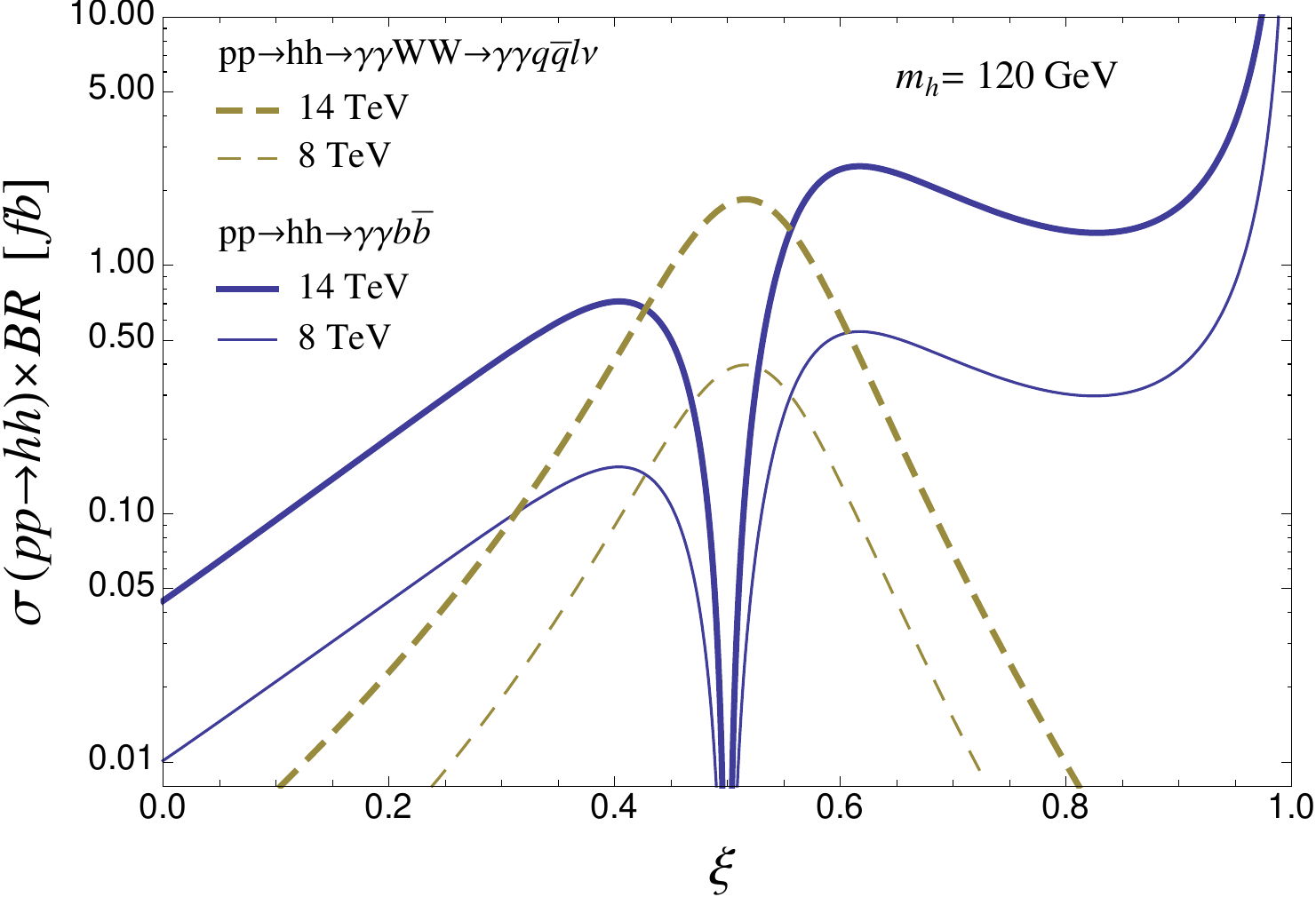} 
\caption{\small
Signal yield per fb$^{-1}$ predicted in the MCHM5 for the two final states $hh\to \gamma\gamma b\bar b$ (solid curves) and 
$hh\to WW\gamma\gamma \to l\nu q\bar q \gamma\gamma$ (dashed curves) as a function of $\xi$. The thick (thin) curves correspond to the
LHC with 14 TeV (8 TeV) center-of-mass energy. The Higgs mass is set to $m_h =120\,$GeV.  The rate has been computed using the
cross section for $gg\to hh$ at LO  in $\alpha_s$  (\textit{i.e.} no $K$-factor is included) given by eq.(\ref{eq:fitxsectot}) and Table~\ref{tab:fit}.
}
\label{fig:xsecxBRMCHM5}
\end{figure}

Figure~\ref{fig:BRs} illustrates how the branching ratios $BR(hh\to \gamma\gamma b\bar b)$ and $BR(hh\to WW\gamma\gamma)$ 
vary with~$c/a$. 
%
%
Strong enhancements compared to the SM prediction are possible for $BR(hh\to WW\gamma\gamma)$ in the fermiophobic limit $c\to 0$.
A fermiophobic composite Higgs can for example arise in the MCHM5  for $\xi\to 1/2$, see eq.~(\ref{eq:par_MCHM5}). 
Although the point $(a=1$, $c=0)$ has been excluded
at 95\%~CL in the range $m_h = 110-192\,$GeV by the combination of all CMS searches~\cite{CMS-PAS-HIG-12-008}, the one predicted by the 
MCHM5 for $\xi=1/2$  $(a=1/\sqrt{2},c=0)$
is still allowed for $m_h \sim 125\,$GeV and in fact could better explain the pattern of observed enhancements in the various $\gamma\gamma$ 
categories of the CMS  analysis~\cite{CMS-PAS-HIG-12-002}.
Figure~\ref{fig:xsecxBRMCHM5} shows the final yield per fb$^{-1}$ predicted in the MCHM5 in the two final states $hh\to \gamma\gamma b\bar b$ and 
$hh\to WW\gamma\gamma \to l\nu q\bar q \gamma\gamma$ as a function of $\xi$ for $m_h = 120\,$GeV. The rate has been computed using the
cross section for $gg\to hh$ at LO  in $\alpha_s$  (\textit{i.e.} no $K$-factor is included) given by eq.(\ref{eq:fitxsectot}) and Table~\ref{tab:fit}. 
%
%
As expected, for small values of $\xi$  the most promising channel is $\gamma\gamma b\bar b$, whose rate can be significantly enhanced compared to the SM expectation.
At 14 TeV, for example, even for $\xi = 0.1$ the signal yield more than doubles. This large sensitivity to small values of $\xi = (v/f)^2$ 
shows that double Higgs production via gluon fusion is an extremely powerful process to probe the Higgs compositeness at the LHC.
Still, the difficulty of isolating the $\gamma\gamma b\bar b$ signal from the background will require large integrated luminosities and will be possible only
in the high-energy phase of the LHC.
In the fortunate situation in which the Higgs is fermiophobic, on the other hand, the enhancement of the $hh\to WW\gamma\gamma \to l\nu q\bar q \gamma\gamma$
final state is so large in the MCHM5 that a first preliminary observation of the signal might be possible at the 8 TeV LHC.
At this energy, for $\xi =1/2$, $m_h = 120\,$GeV  and 20 fb$^{-1}$ of integrated luminosity the MCHM5 predicts $\sim 15$ signal events before cuts.~\footnote{We
included a $K\text{-factor}=2$ in the estimate, which is the value obtained in Ref.~\cite{Dawson:1998py} at 14 TeV, assuming that a similar result also applies at 8 TeV.
For $m_h =125\,$GeV the number of signal event is $\sim 10$.}
Considering that the SM background is expected to be rather small for a final state with two photons and one isolated lepton, this number of events might be sufficient
to establish the observation of the signal.
For the same value of c.o.m. energy and integrated luminosity, the MCHM5 predicts  $\sim 42$ and $\sim 27$ signal events (before cuts) respectively in 
$hh\to 4W\to l^{\pm} l^{\pm} \nu\nu 4q$ (two same-sign leptons) and $hh\to 4W\to 3l 3\nu q\bar q$. These high rates suggest that 
it might be possible to distinguish the $hh\to 4W$ signal over the SM background even at 8 TeV.

The results discussed in this section are rather encouraging, and show that double Higgs production can be an important process to extract or constrain
the $t\bar t hh$ interaction and, more in general, to probe the Higgs compositeness. 
However, a more robust assessment of the LHC sensitivity in this sense requires a dedicated analysis of each of the relevant final states and a careful estimate of the 
background. In the next section we will focus on the $\gamma\gamma b\bar b$ channel and use the studies of Ref.~\cite{Baur:2003gp} to get a first determination of the
precision which can be obtained on $c_2$ at the 14 TeV~LHC.

\section{Analysis of the $b\overline b\gamma\gamma$ channel}
\label{sec:bbgammagamma}

The analysis of the $gg \rightarrow hh \rightarrow b \overline b \gamma \gamma$ process
performed in Ref.~\cite{Baur:2003gp} aimed at measuring the 
Higgs trilinear coupling, and assumed
SM values for the other couplings. In this section we make use of the results of~\cite{Baur:2003gp} to estimate the LHC sensitivity on the
$t\bar t hh$ non-linear interaction.

We assume a center-of-mass energy $\sqrt{s} =14\,$TeV and set $m_h = 120\ {\rm GeV}$. 
Given the value of the signal rate and the large SM background, an analysis of $b\bar b \gamma\gamma$ at 8 TeV seems rather challenging 
and for this reason it will not be considered here.
We expect our results at $14\,$TeV to be representative of what the LHC sensitivity will be in its future high-energy phase,
even if the actual value of c.o.m. energy turns out to be different (on the prospects of 14 and 13 TeV see for example~\cite{chamonix}).
The analysis is performed at the parton level: signal events are generated by means of our implementation of double Higgs production via gluon fusion
in ALPGEN, while the computation of the background processes is taken from~\cite{Baur:2003gp}.
We take the NLO QCD corrections to $gg\to hh$ into account by multiplying the LO cross section by a factor $K=2$~\cite{Dawson:1998py},~\footnote{ Notice 
that the authors of Ref.~\cite{Baur:2003gp}  used a factor $K=1.65$ for the LHC at 14 TeV, which is the one appropriate for their choice of the renormalization 
and factorization scale  $Q=m_h$. As previously discussed, we set instead $Q = m(hh)$.}
while we  neglect the smaller contribution of the vector-boson-fusion process to double Higgs production. To ensure an effective
suppression of the background, we select events with two photons
and two $b$-jets. Two photon tags are necessary to suppress the  huge QCD background. On the other hand, the optimal
number of  $b$-tags depends on the strength of the signal and on the collider energy.
In Ref.~\cite{Baur:2003gp} only one $b$-tag was required for the analysis at the 14 TeV~LHC.
This choice was motivated by the necessity of preserving the small SM
signal as much as possible, at the expense of having to cope with a larger background.
In our case the signal cross section is much larger than the SM one in a sizable part of the parameter space.
Furthermore, recent studies have shown that a larger $b$-tagging efficiency than that assumed in Ref.~\cite{Baur:2003gp}
is possible at the LHC~\cite{ATLAS_btag}, 
maintaining an acceptable rejection rate on jets.
In particular, we adopt the following conservative estimates for the efficiencies ($\epsilon$) 
and fake rates ($r$) for $b$-tagging~\cite{ATLAS_btag} and photon reconstruction~\cite{ATLAS_gamma}:~\footnote{ The authors of Ref.~\cite{Baur:2003gp} 
use instead $\epsilon_b = 0.5$ and fake rates
$r_{c\rightarrow b} = 13$ and $r_{j\rightarrow b} = 140$.}
\begin{equation}
\label{eq:taggingeff}
\epsilon_\gamma = 0.8\,, \qquad r_\gamma = 2500\,, \qquad
\rule{0pt}{1.25em}\epsilon_b = 0.7\,, \qquad r_{c\rightarrow b} = 5\,,
\qquad r_{j\rightarrow b} = 25\,.
\end{equation}
This allows us to require two $b$-tags, thus obtaining a stronger suppression
of the background at the price of an affordable reduction of the signal. 
We find that including the additional category of events with one $b$-tag in the analysis does not sensibly improve our results.

Following~\cite{Baur:2003gp}, we impose the set of kinematic cuts
\begin{equation} \label{eq:acceptance_cuts}
\begin{gathered}
p_T(b) > 45\ {\rm GeV}\,, \qquad  |\eta(b)| < 2.5\,,   \qquad \Delta R(b,b) > 0.4\,,\\[0.15cm]
  m_h - 20\ {\rm GeV} < m({b \overline b}) < m_h + 20\ {\rm GeV}\,,  \\[0.25cm]
p_T(\gamma) > 20\ {\rm GeV}\,, \qquad |\eta(\gamma)| < 2.5\,, \qquad \Delta R(\gamma,\gamma) > 0.4\,,\\[0.15cm]
 m_h - 2.3\ {\rm GeV} < m({\gamma \gamma}) < m_h + 2.3\ {\rm GeV}\,, \\[0.25cm]
\Delta R(\gamma, b) > 0.4\, ,
\end{gathered}
\end{equation}
which have a high selection efficiency on the signal and ensure that the $b\overline b$ and $\gamma \gamma$  invariant masses
are reconstructed in the given windows around the Higgs mass. To take into account the detector resolution, we
have assumed a $79\%$ efficiency for the reconstruction of the $b\overline b$ pair and a $79\%$ efficiency for the reconstruction of the $\gamma\gamma$ pair 
in the signal, as done in Ref.~\cite{Baur:2003gp}.
After the above cuts, the most important irreducible backgrounds come from the  $b\overline b\gamma\gamma$ continuum,
and potentially by single Higgs production in association with two $b$-quarks ($h(\rightarrow \gamma\gamma) b\overline b$)
or two photons ($h(\rightarrow b\overline b)\gamma\gamma$). The reducible backgrounds
are QCD processes ($c\bar c \gamma\gamma$, $b\bar b \gamma j$, $c\bar c\gamma j$, $b\bar b jj$, $c\bar c jj$, $\gamma\gamma jj$, $\gamma jjj$, $jjjj$) or single-Higgs production processes ($hjj$ and $hj\gamma$) 
where some of the jets fake one or more $b$-quarks or photons.
While the QCD backgrounds are universal and as such are not affected by New Physics, those
coming from single Higgs production depend on the value of the modified Higgs couplings. 
In the SM, all single Higgs processes are much smaller than the double Higgs signal after the cuts of eq.(\ref{eq:acceptance_cuts}), and
can be safely neglected~\cite{Baur:2003gp}. This approximation is still valid in our
context, and for this reason we will not include these backgrounds in our analysis.
The list of relevant processes and their cross sections after the cuts of eq.~(\ref{eq:acceptance_cuts}) (without including $b$ and photon reconstruction efficiencies)
is reported in Table~\ref{table:backgrounds}.
%
\begin{table}
\centering
\begin{tabular}{|c|c|c|c|c|c|c|c|c|c|}
\hline
\hline
\rule{0pt}{1.1em} \!\!used cuts\!\! & $b\overline b \gamma\gamma$ & $c\overline c \gamma\gamma$ &
$b\overline b \gamma j$ & $c\overline c \gamma j$ & $jj \gamma\gamma$ & $b\overline b jj$ &
$c\overline c jj$ & $\gamma jjj$ & $jjjj$\\
\hline
\rule{0pt}{1.1em} \!\!eq.~(\ref{eq:acceptance_cuts})\!\! & $0.056$ & $0.42$ & $65$ & $250$ & $11$ & $2.5\!\times\! 10^4$
& $2.5\!\times\! 10^4\!$ & $7700$ & $5\!\times\! 10^6$\\
\rule{0pt}{1.1em} \!\!+ eq.~(\ref{eq:background_cuts})\!\! & $0.0060$ & $0.0215$ & $8.28$
& $17.0$ & $0.84$ & $4520$ & $4520$ & $364$ & $4\!\times\! 10^5$ \\
\hline
\hline
\rule{0pt}{1.1em} \!\!+ tags\!\! & $\!0.0019\!$ & $\!5\!\times\! 10^{-4}\!$
& $\!0.0013\!$ & $\!2\!\times\! 10^{-4}\!$ & $\!9\!\times\! 10^{-4}\!$
& $\!4\!\times\! 10^{-4}\!\!$ & $\!3\!\times\! 10^{-5}\!\!$ & $\!2\!\times\! 10^{-4}\!$ & $\!1\!\times\! 10^{-4}\!$\\
\hline
\hline
\end{tabular}
\caption{\small
Cross sections (in ${\rm fb}$) of the main QCD backgrounds to
$hh \rightarrow b\overline b \gamma\gamma$ at the 14 TeV~LHC.
The values of the cross sections after the cuts of eqs.~(\ref{eq:acceptance_cuts})
and (\ref{eq:background_cuts}) are taken from \cite{Baur:2003gp}, and include an additional rescaling factor 1.3 
introduced to take into account the increase that can possibly come from NLO corrections. The last line reports the value
of the cross sections after the inclusion of the $b$-jet and photon tagging efficiencies of eq.~(\ref{eq:taggingeff}).
}
\label{table:backgrounds}
\end{table}

A further suppression of the background can be obtained by exploiting the particular topology of the signal, where the two Higgs bosons are produced
back to back in the center-of-mass frame. We thus select events where the $\gamma\gamma$ pair has a small
opening angle, while the minimal angular separation between a $b$-jet and a photon is large. We require~\cite{Baur:2003gp}:
\begin{equation}\label{eq:background_cuts}
\Delta R(\gamma, b) > 1.0\,, \qquad \Delta R(\gamma, \gamma) < 2.0\,.
\end{equation}
In most of the parameter space these cuts imply a moderate reduction of the signal ($20\% - 40\%$), while
the total background is suppressed by one order of magnitude. The corresponding background cross sections are reported in
the second line of Table~\ref{table:backgrounds}.

Finally, the reducible backgrounds are drastically suppressed once the efficiencies for reconstructing
two photons and two $b$-jets of eq.(\ref{eq:taggingeff}) are included.
The resulting final cross sections are shown in the last line of Table~\ref{table:backgrounds}.
The corresponding total background cross section is $r_b = 5.5\ {\rm ab}$.

After performing the kinematic cuts of eqs.~(\ref{eq:acceptance_cuts}), (\ref{eq:background_cuts}) and  including the efficiencies for the 
reconstruction of the $\gamma\gamma$ and $b\bar b$ pairs, the tagging efficiencies (\ref{eq:taggingeff}), and the $K$-factor,
the signal rate ($r_s \equiv \sigma(pp\to hh)\times BR(hh\to \gamma\gamma b\bar b)$) at $m_h = 120\ {\rm GeV}$ is well approximated by the formula
\begin{equation} \label{eq:approximate_xsection_cuts}
\begin{split}
r_s =\frac{BR(hh\to \gamma\gamma b\bar b)}{BR(hh\to \gamma\gamma b\bar b)_{SM}} \times (49.3\,\text{ab}) \Big[ 
& c_2^2+\left(0.407\, c^2\right)^2 + \left(0.101\, cd_3\right)^2 -1.76\, c_2\left(0.407\, c^2\right) \\
& -1.82 \left(0.407\, c^2\right)\left(0.101\, cd_3\right)+1.72\, c_2\left(0.101\, cd_3\right)  \Big]\, .
\end{split}
\end{equation}
For the SM case ($c=d_3=1$, $c_2=0$) we find $r_s = 4.9\ {\rm ab}$.
Notice that, compared to the fit of eq.(\ref{eq:fitxsectot}) at 14~TeV (see Table~\ref{tab:fit}),
the cuts have further weakened the dependence of the cross section on $d_3$, since their
efficiency is smaller for events with low $m(hh)$ invariant mass.

\subsection{Results}

Using the signal and background rates derived above, we can estimate the sensitivity of the 14~TeV~LHC on $pp\to hh \to b\bar b\gamma\gamma$ for $m_h = 120\,$GeV.
Since large luminosities are typically needed to distinguish the signal from SM background, we expect that
by the time the analysis of double Higgs production is performed, the Higgs branching ratios to $\gamma\gamma$ and $b\bar b$ and the
linear couplings $a$, $c$ are known with good accuracy. Double Higgs production can thus be used to extract (or set limits on) the couplings $c_2$ and $d_3$.

\begin{figure}
\centering
\includegraphics[width=.445\textwidth]{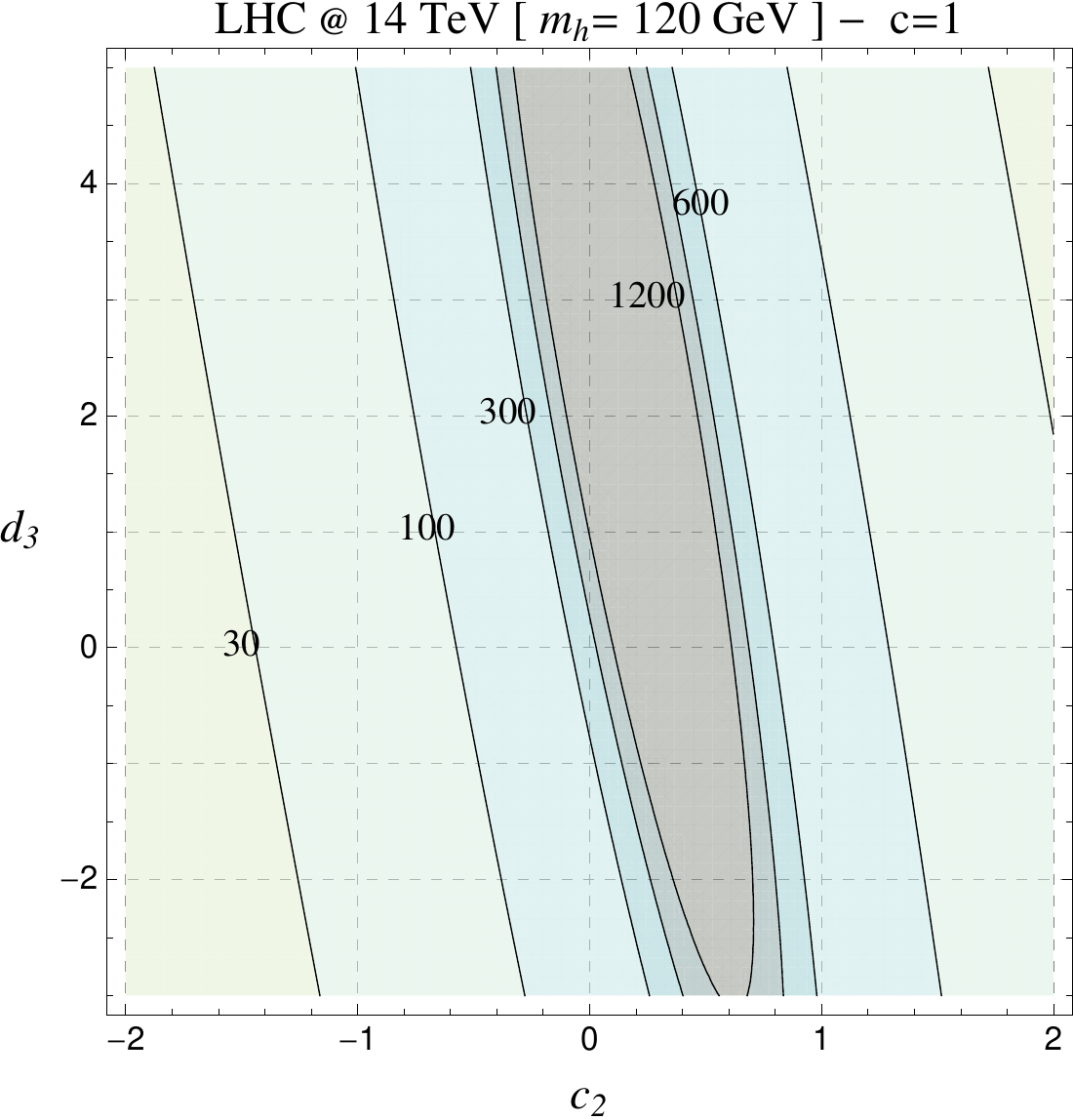}
\hspace{1.5em}
\includegraphics[width=.44\textwidth]{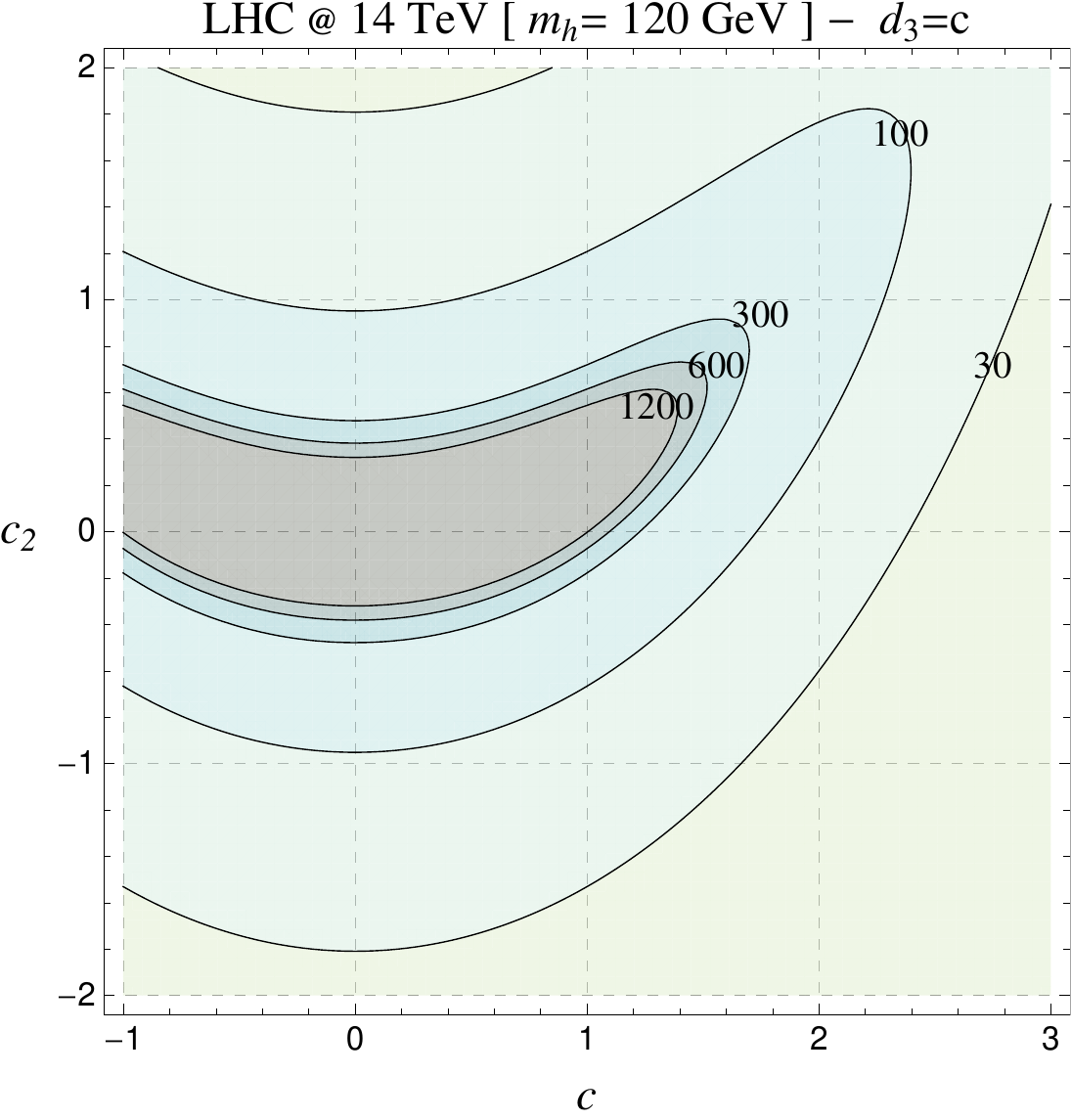}
\caption{\small
Isocurves of discovery luminosity (in fb$^{-1}$) at the 14 TeV LHC in the plane $(c_2, d_3)$ for $c=1$ (on the left) and in the plane $(c, c_2)$ for $d_3 =c$ (on the right).
Outside each contour, the $pp\to hh\to \gamma\gamma b\bar b$ signal can be discovered with the corresponding integrated luminosity.
In both plots the Higgs mass is set to $m_h =120\,$GeV and the Higgs decay branching ratios are fixed to their SM values. See the Appendix for the
definition of discovery luminosity.
}
\label{fig:discoveryplots}
\end{figure}

\begin{figure}
\centering
\includegraphics[width=0.55\textwidth]{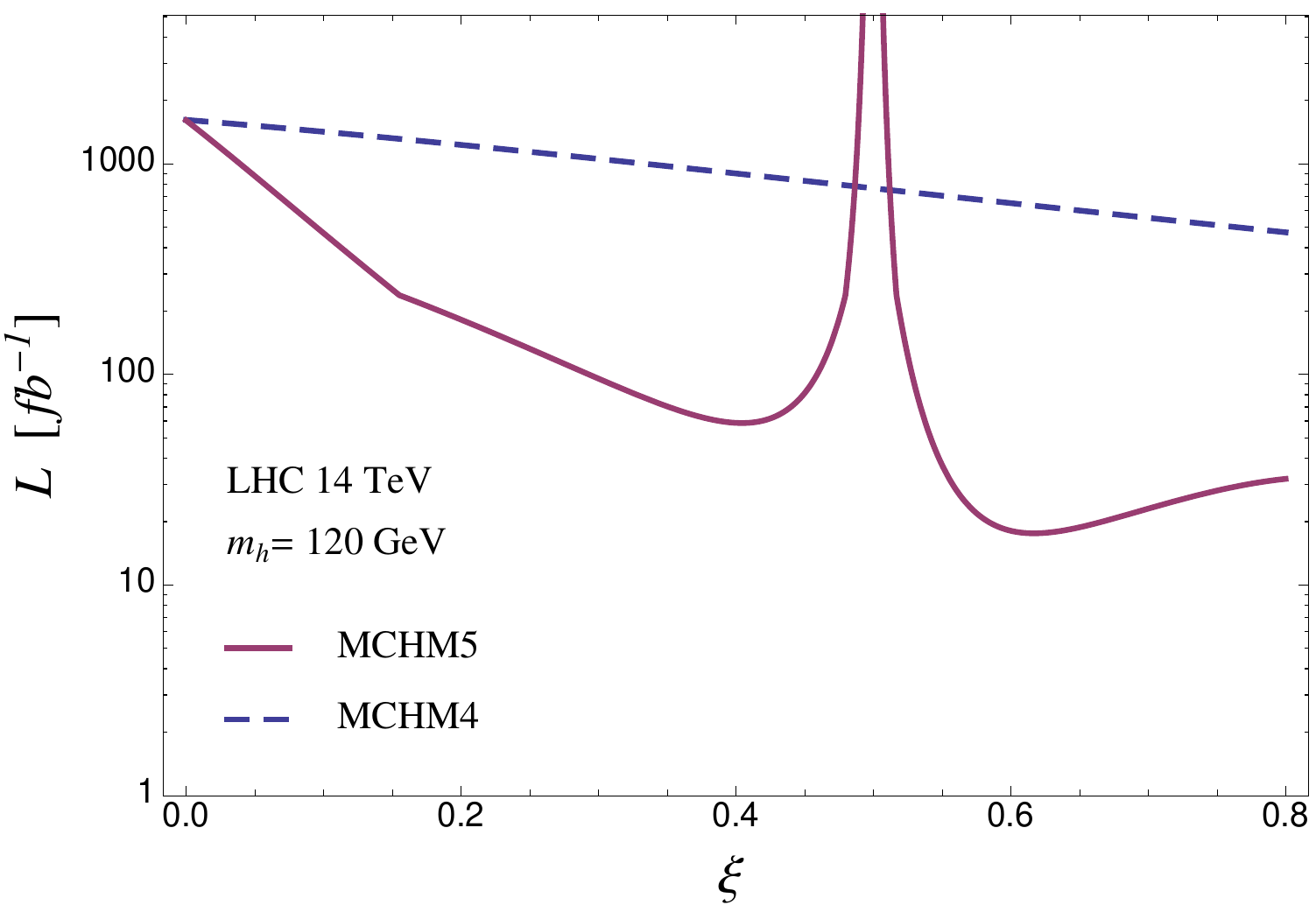}
\caption{\small
Discovery luminosity at the 14 TeV LHC in the MCHM4 (dashed blue curve) and MCHM5 (continuous purple curve) as a function of $\xi$.
The Higgs mass is set to $m_h =120\,$GeV. See the Appendix for the definition of discovery luminosity.
}
\label{fig:discoveryMCHM}
\end{figure}

Figure~\ref{fig:discoveryplots} shows the luminosity required to discover the signal
as a function of $c$, $c_2$ and $d_3$, assuming that the  branching ratio BR($hh\to \gamma\gamma b\bar b)$ has the value predicted in the SM.~\footnote{The definition 
of discovery luminosity and the details of our statistical analysis are discussed in the Appendix.} 
%
%
The plots on the left and on the right show the luminosity contours respectively in the plane $(c_2, d_3)$ for $c=1$, and in the plane $(c, c_2)$ for $d_3 =c$.
As expected, the sensitivity on $c$ and $c_2$ is  stronger than that on the Higgs trilinear coupling $d_3$.
In particular, while a discovery in the SM would require at least $1200\,\text{fb}^{-1}$, we find that much lower luminosities are sufficient even for moderately 
small values of~$c_2$. 
Figure~\ref{fig:discoveryMCHM} shows the corresponding discovery luminosity in the composite Higgs models MCHM4 and MCHM5 as a function of $\xi$.
%
%
We find that values of $\xi$ as small as $0.15$ can be probed with $300\,\text{fb}^{-1}$ of integrated luminosity.
Compared to other processes like double Higgs production via vector boson fusion~\cite{Contino:2010mh}, these results show
that $gg\to hh$ can be extremely powerful to study the non-linear couplings of a composite Higgs and thus  probe its strong
interactions.

\begin{figure}
\centering
\includegraphics[width=.464\textwidth]{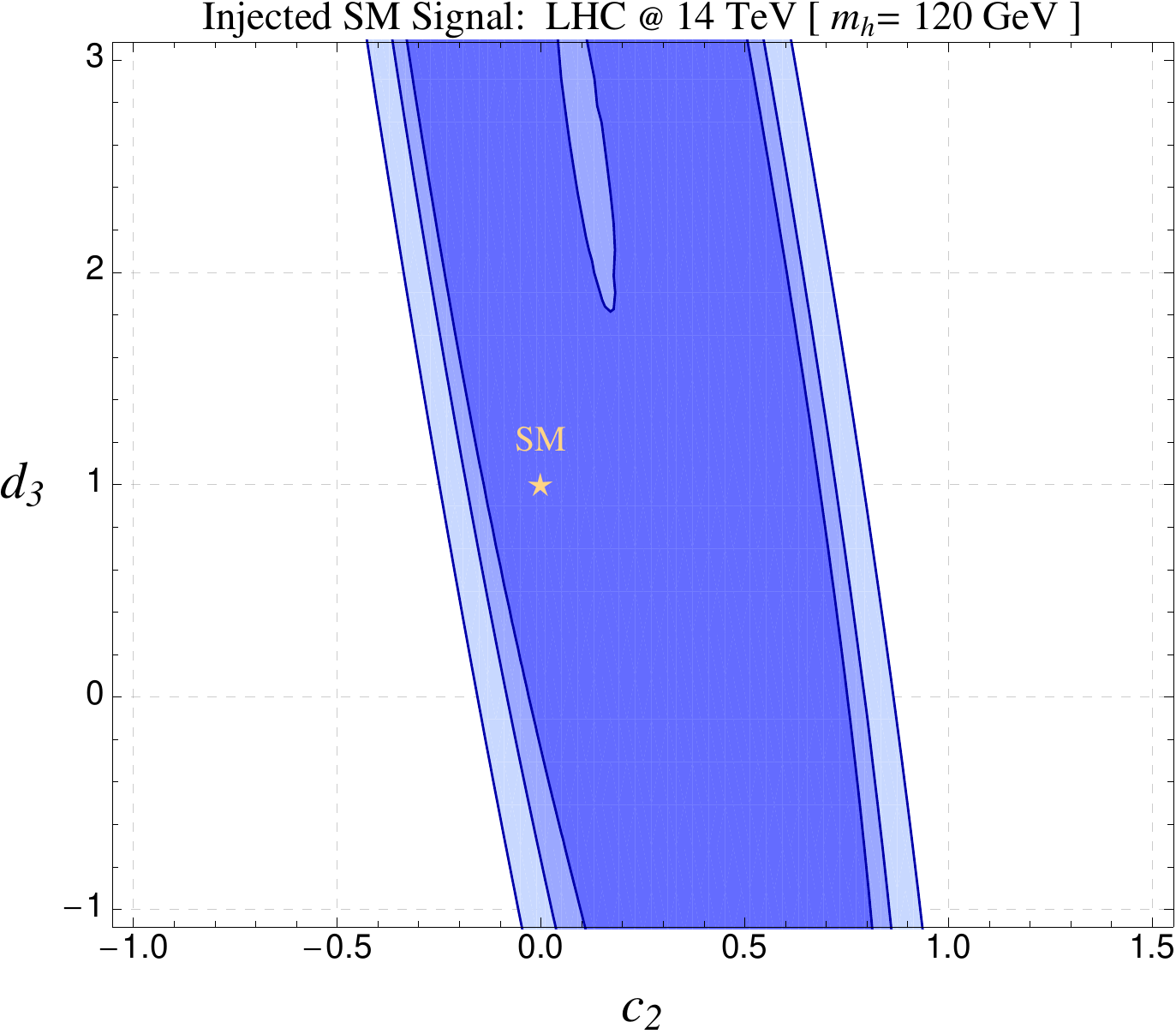}
\hspace{0.5em}
\includegraphics[width=.46\textwidth]{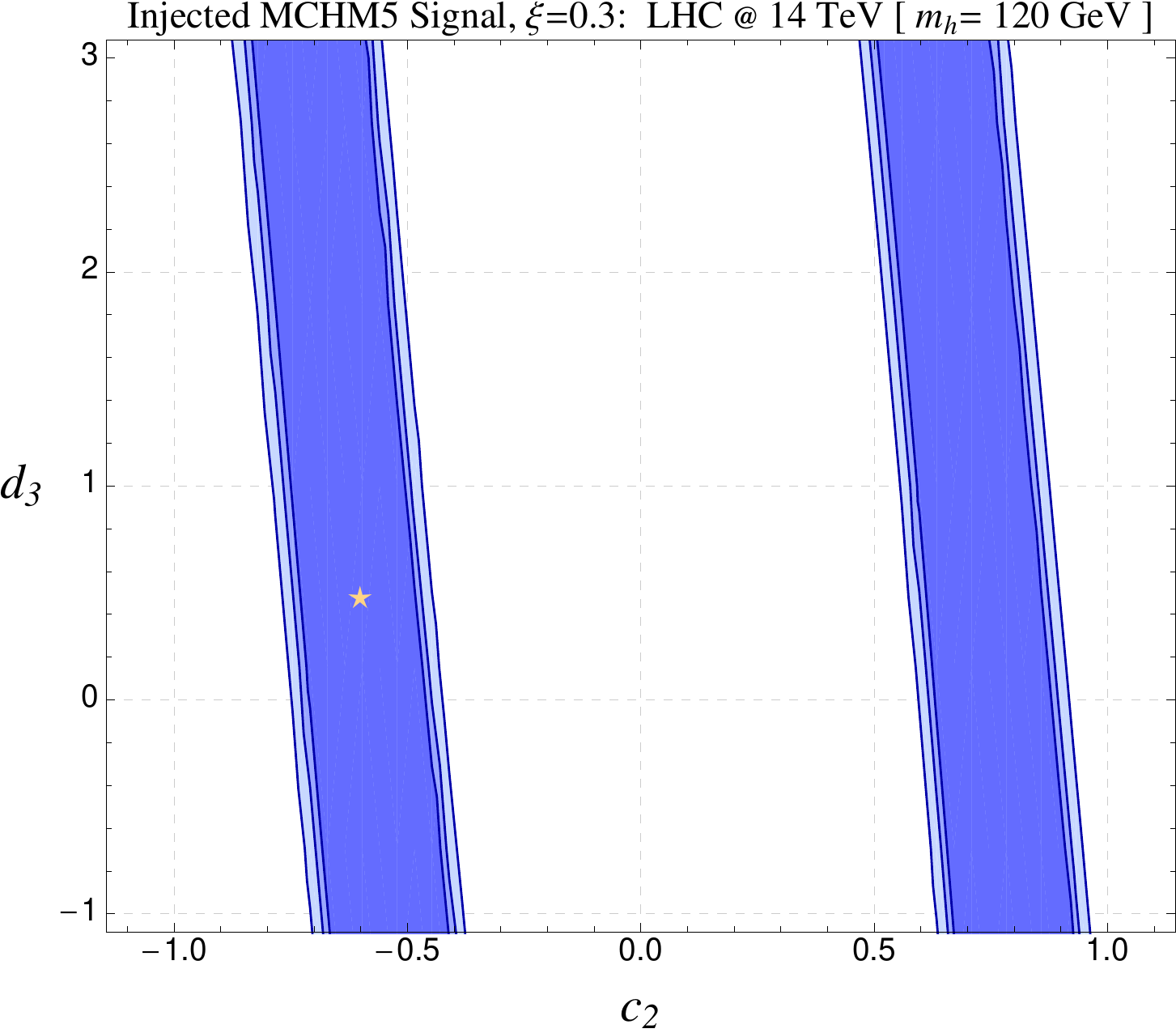}
\caption{\small
Regions of 68\% probability in the plane $(c_2,d_3)$ obtained with 300 (light blue area), 600 (medium dark blue area) and 1200~fb$^{-1}$ 
(darker blue area) of integrated luminosity. On the left: injected signal is the SM  ($c=d_3=1$, $c_2=0$); On the right: injected signal is the MCHM5 with $\xi=0.3$
 ($c=d_3 = 0.48$, $c_2 = -0.6$). Both plots are obtained by assuming that the branching fraction $BR(hh\to \gamma\gamma b\bar b)$ and the coupling $c$ 
have been measured from single-Higgs processes with a 20\% uncertainty.
}
\label{fig:precision}
\end{figure}

\begin{figure}
\centering
\includegraphics[width=0.5\textwidth]{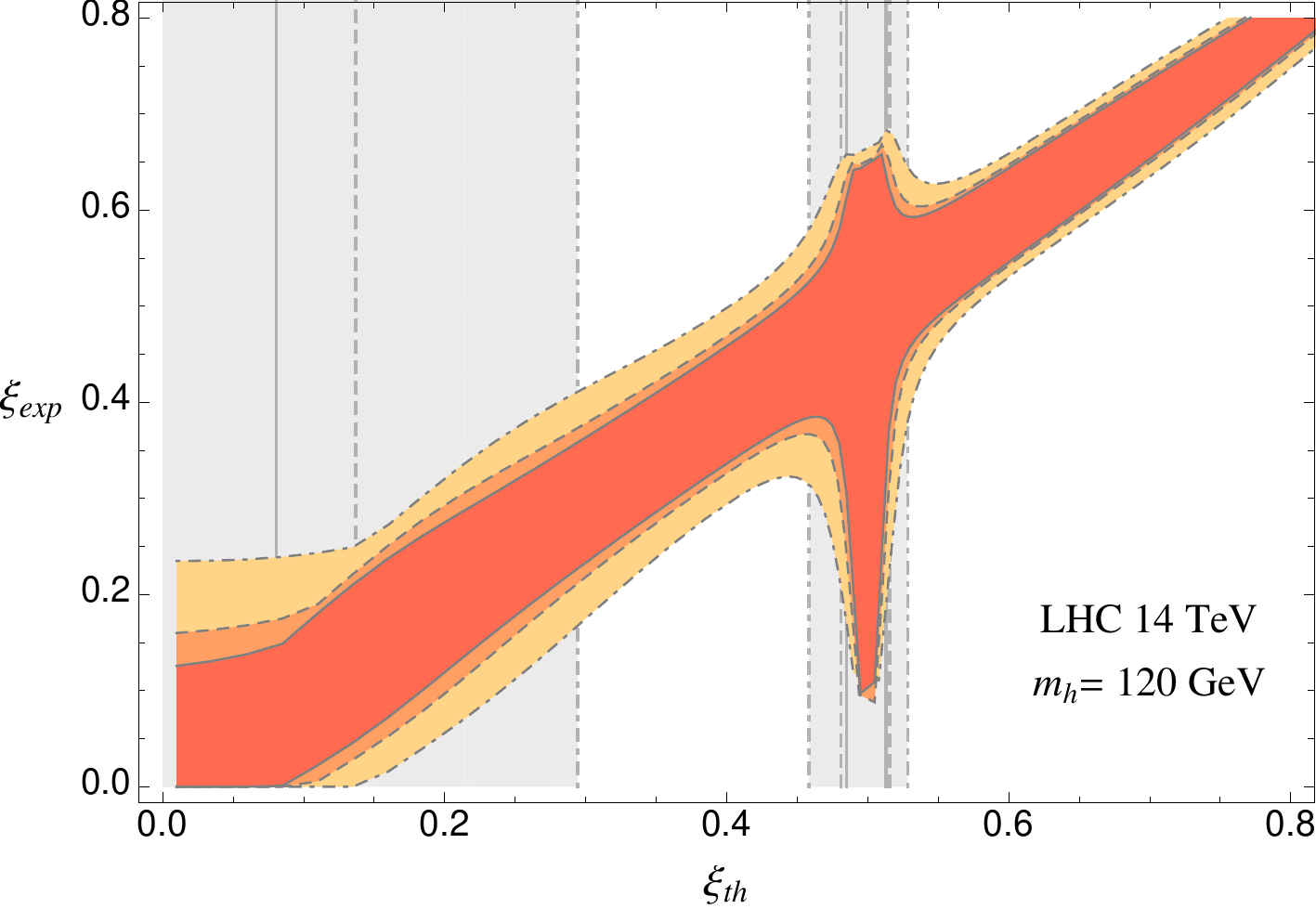}
\caption{\small
Precision on the parameter $\xi$ in the MCHM5 which can be obtained at the 14 TeV LHC for $m_h =120\,$GeV from the analysis of  
$pp\to hh \to \gamma\gamma b\bar b$. 
For each injected value $\xi_{th}$, the solid, dashed and dot-dashed curves (red, orange and yellow regions) show the 68\% probability interval which is expected on 
the measured  value $\xi_{exp}$ respectively with 600, 300 and 100~fb$^{-1}$.  The gray regions delimited by the solid, dashed and dot-dashed vertical lines indicate
the values of $\xi$, respectively with 600, 300 and 100~fb$^{-1}$, for which the signal rate is too small to make a discovery, (see Fig.~\ref{fig:discoveryMCHM}).
In particular, for $\xi \to 0.5$ the $h\to b\bar b$ decay rate vanishes (the Higgs becomes fermiophobic), and the signal cannot be distinguished from the SM 
background.
The curves are derived by assuming that the branching fraction $BR(hh\to \gamma\gamma b\bar b)$ is determined with 20\%
accuracy from single-Higgs production.
}
\label{fig:precisionMCHM}
\end{figure}

Once a discovery is established, one can measure the couplings $c_2$ and $d_3$ by using the value of $c$ and of the Higgs branching ratios
determined in single-Higgs processes.
The left plot of Fig.~\ref{fig:precision} shows the region of $68\%$ probability in the plane $(c_2, d_3)$ with 300, 600 and 1200~fb$^{-1}$ (light, medium 
and dark blue regions) obtained by injecting the SM signal ($c=d_3=1$, $c_2=0$) and assuming that the coupling $c$ and the branching fraction
$BR(hh\to \gamma\gamma b\bar b)$ have been determined with a 20\% accuracy.~\footnote{That is: the rate of observed 
events  is assumed to be that predicted in the SM with $m_h =120\,$GeV. The uncertainties on $c$ and $BR(hh\to \gamma\gamma b\bar b)$ have been
taken into account by marginalizing the 2-dimensional likelihood over two nuisance parameters, see Appendix.}
%
%
In this case the precision on $c_2$ is poor even with 1200~fb$^{-1}$, while $d_3$ is basically unconstrained.
A more precise determination of $c_2$ can be obtained if its value is non-vanishing. The right plot of Fig.~\ref{fig:precision} shows
the case in which the injected signal is that of the MCHM5 with $\xi=0.3$, corresponding to ($c=d_3 = 0.48$, $c_2 = -0.6$).
It assumes that the branching fraction $BR(hh\to \gamma\gamma b\bar b)$ and the coupling $c=0.48$ predicted by this model have been measured with 
20\% accuracy in single-Higgs processes.
We find that with 300~fb$^{-1}$ the coupling $c_2$ can be determined, up to a  discrete ambiguity, with a precision of  $\sim 20-30\%$.~\footnote{This
improves to $\sim 15-20\%$ if the uncertainty on $c$ and $BR(hh\to \gamma\gamma b\bar b)$ is negligible.}
On the other hand, even in this case $d_3$  remains largely unconstrained with our analysis.
Finally, Fig.~\ref{fig:precisionMCHM} shows how precisely the parameter $\xi$ can be determined in the MCHM5 through $gg\to hh \to \gamma\gamma b\bar b$ 
by making use only of the value of the decay branching ratios determined in single-Higgs processes (that is: without fixing $c$ to its measured value in the fit). 
As before, we assume that the branching fraction $BR(hh\to \gamma\gamma b\bar b)$ is known with an error of 20\%.
%
%
For each injected value $\xi_{th}$, the solid, dashed and dot-dashed curves (red, orange and yellow regions) show the 68\% probability interval which is expected on 
the measured  value $\xi_{exp}$ respectively with 600, 300 and 100~fb$^{-1}$. 
For example, with 300~fb$^{-1}$ of integrated luminosity, $\xi=0.2$ can be measured with a precision of $\sim 45\%$.
The solid, dashed and dot-dashed vertical lines indicate the range of  $\xi$, respectively for 600, 300 and 100~fb$^{-1}$, for which the expected signal yield 
is sufficiently large to establish a discovery (see Fig.~\ref{fig:discoveryMCHM}).

\section{Conclusions}

The discovery of a light Higgs-like scalar at the LHC will mark a first important step forward in our comprehension
of the  electroweak symmetry breaking mechanism. 
Precise knowledge of the strength of its interactions with SM fields can shed light on the origin of the Higgs boson
and indicate if the new dynamics at the electroweak scale is weakly or strongly interacting.
New strong dynamics can form the Higgs as a bound state and solve naturally the hierarchy problem of the Standard Model.
In this case the Higgs boson itself interacts strongly at large energies due to its modified linear couplings to SM fields
and the existence of new non-linear  interactions. 

In this paper we have performed a first model-independent study of double Higgs production via gluon fusion, $gg\to hh$, and
we have shown that  its cross section  is greatly enhanced by the non-linear  interaction $t\bar t hh$.
Such new vertex gives a contribution to the scattering amplitude of $t\bar t \to hh$ that grows with the energy, ${\cal A}(t\bar t\to hh) \sim (E m_t)/v^2$,
and as such it is a genuine signature of the underlying strong dynamics. In the process $gg\to hh$ the $t\bar t hh$ vertex mediates 
a new diagram containing a top-quark loop which grows logarithmically at high energies and does not lead to a violation of perturbative unitarity 
(see eq.(\ref{eq:triangle_c2_behaviour})).
However, it does lead to a strong numerical enhancement of the cross section compared to the SM, as first noticed by the authors of Ref.~\cite{Grober:2010yv} 
in the context of  the MCHM composite Higgs models.  The origin of this enhancement 
can in part be traced back to the sizable destructive interference which occurs in the Standard Model between
the box and the triangle diagram with Higgs exchange. A similar cancellation has been found to take place at large $\hat s$ in  $gg\to ZZ$~\cite{Glover:1988fe} 
and $gg\to WW$~\cite{Accomando:2007xc}, 
and interpreted as  a relic of the cancellation dictated by unitarity between the energy-growing amplitudes in the sub-process $t\bar t\to VV$.
On the other hand, in the case of double Higgs production the cancellation is in the \textit{total} cross section (\textit{i.e.} not necessarily at large $\hat s$). Furthermore,
none of the diagrams which contribute to  $t\bar t\to hh$ in the SM grows with the energy, so that the cancellation between
the box and the  triangle diagram in $gg\to hh$ should be rather seen as a numerical accident, not driven by the unitarity of its subprocess. 

The strong enhancement of the $gg\to hh$ cross section makes this process quite powerful to measure or constrain the strength of the 
$t\bar t hh$ interaction, which we have denoted as $c_2$. This should be compared with the much weaker sensitivity on the Higgs trilinear
interaction, $d_3$, which makes the extraction of this latter coupling extremely challenging at the LHC.
In particular, the weaker  dependence on $d_3$ follows from an extra suppressing factor $(m_h^2/\hat s)$ carried by the triangle diagram
with Higgs exchange,  which thus contributes mainly at threshold.
In order to assess the LHC precision on $c_2$ we have made use of the results of Refs.~\cite{Baur:2002rb,Baur:2002qd,Baur:2003gpa,Baur:2003gp}, where 
double Higgs production via gluon fusion was studied in the context of the SM and several Higgs decay channels were investigated.
In a generic scenario of New Physics, what is the best final state largely depends  on the value of the Higgs decay branching ratios.
In particular, enhanced branching ratios can combine with the  increase in the double Higgs production cross section and  lead to dramatic effects at the LHC.
For example, in the (fermiophobic) limit in which the linear couplings of the Higgs to the SM fermions are suppressed ($c\to 0$), the $h\to WW$
and $h\to \gamma\gamma$ branching ratios can be sensibly enhanced compared to their SM values. 
In this case the final states $hh\to WWWW$ and $hh\to WW\gamma\gamma$ seem
to be extremely promising, and  might  be visible even at the 8 TeV LHC.   
At this energy  and with $L = 20\,$fb$^{-1}$, for example, we find that for $m_h = 120\,$GeV  the MCHM5 at $\xi =0.5$
predicts $\sim 15$ signal events in $hh\to WW\gamma\gamma\to l\nu q\bar q \gamma\gamma$ (see Fig.~\ref{fig:xsecxBRMCHM5}).

If the Higgs decay branching ratios  do not differ much from their SM values and the Higgs boson is light,
the most powerful final state should be $hh\to b\bar b\gamma\gamma$, as suggested by the analysis of Ref.~\cite{Baur:2003gp}. 
Even in this case, the signal rate can be significantly enhanced compared to the SM prediction if $c_2$ is not too small.
At 14 TeV, for example, the signal rate predicted in the MCHM5 with $\xi =0.1$ is larger than the SM one by more than a factor two.
This sensitivity on small values of $\xi = (v/f)^2$ shows that double Higgs production via gluon fusion is an extremely powerful process 
to probe the Higgs compositeness at the LHC.
In order to estimate the LHC sensitivity on $c_2$ in a model-independent way, we
have followed the strategy proposed in Ref.~\cite{Baur:2003gp} and performed a Montecarlo study of $pp\to hh\to b\bar b\gamma\gamma$.
We have computed the SM background cross section by using the results reported in \cite{Baur:2003gp} and rescaling them to take account of 
updated $b$ and $\gamma$  efficiencies and rejection factors.
The results that we obtained are quite encouraging. With $L =300\,\text{fb}^{-1}$ the 14 TeV LHC can probe values $c_2 \lesssim -0.2$ and
$c_2 \gtrsim 0.8$ if $c, d_3\sim 1$
(see Fig.~\ref{fig:discoveryplots}). In the case of the MCHM5, an integrated luminosity of $300\,\text{fb}^{-1}$ is sufficient to discover a signal with $\xi \gtrsim 0.15$ 
(see Fig.~\ref{fig:discoveryMCHM}).
In general, once the signal can be statistically distinguished from the background and a discovery is made, the value of $c_2$ can be extracted with good accuracy.
For example, we find that by injecting a signal with $c = d_3 = 0.48$ and $c_2=-0.6$ (as predicted in the MCHM5 for $\xi=0.3$), the coupling $c_2$ can be measured,
up to a discrete ambiguity, with a precision of $\sim 20-30\%$ (see Fig.~\ref{fig:precision}). 

\enlargethispage{0.5cm}

Our partonic analysis of $pp\to hh \to b\bar b\gamma\gamma$ should be considered as a first estimate of the LHC potentiality, although
we expect it to be robust and moderately conservative. For example, we have followed a cut-based strategy to reduce the background, although 
a realistic analysis will certainly make use of shape variables and  extract the background from data, similarly to what has been done 
by the ATLAS and CMS collaborations in single Higgs searches. Also notice that we did not make use of the information on the total invariant mass
distribution, $d\sigma/d m(b\bar b\gamma\gamma)$, which was instead used in Ref.~\cite{Baur:2003gp} to further increase the signal significance.
Finally, a more precise assessment of the LHC sensitivity  will  require full inclusion of showering and hadronization effects, as well as a detector simulation.

\section*{Acknowledgments}

We would like to thank A.~Azatov, A. Lazopoulos and R. Rattazzi for useful discussions.
G.P. and A.W. thank the physics department of the University of Rome ``La Sapienza'' for hospitality
during the completion of this work. 
F.P. and M.M. would like to thank the CERN TH-Unit for partial support and hospitality during
several stages of the work.
A.W. and R.C. were partly supported by the ERC Advanced Grant No.~267985 
{\em Electroweak Symmetry Breaking, Flavour and  Dark Matter: One Solution for Three Mysteries (DaMeSyFla)}.
The work of A.W. was supported in part by the European Programme Unification
in the LHC Era, contract PITN-GA-2009-237920 (UNILHC).
The work of F.P. and M.M. was supported
in part by the Research Executive Agency (REA)
of the European Union under the Grant Agreement
number PITN-GA-2010-264564
(LHCPhenoNet)

\appendix
\section{Appendix}
\enlargethispage{0.5cm}

We report here the details of the statistical analysis used to derive our results. 
We follow  the Bayesian approach~\footnote{See for example Ref.~\cite{D'Agostini:2003nk} for a primer.}
and construct a posterior probability for the total event rate $r$, 
\begin{equation}
p_{\cal L}(r | N) \propto L(N | r {\cal L})\, \pi(r)
\end{equation}
for given number of observed events $N$ and luminosity ${\cal L}$.
We denote with $\pi(r)$ the prior distribution and with $L(N | r {\cal L})$ the likelihood function, which we take to be a Poisson distribution
\begin{equation}
L(N | r {\cal L}) = \frac{\displaystyle e^{- r {\cal L}} \, (r {\cal L})^N }{N!}\, .
\end{equation}

For the plots of Figs.~\ref{fig:discoveryplots}, \ref{fig:discoveryMCHM}
we use a prior distribution which is flat for $r >0$ and vanishing otherwise, and normalize the probability so that $\int_0^\infty \! d r\, p_{\cal L}(r | N) =1$.
The discovery contours of Figs.~\ref{fig:discoveryplots}  and \ref{fig:discoveryMCHM} are obtained by setting the number of observed events
to the total value $N = (r_s + r_b) {\cal L}$ expected in each point of the Higgs couplings' parameter space:  $r_s = r_s(a,c,c_2,d_3)$ for Fig.~\ref{fig:discoveryplots}, 
$r_s = r_s(\xi)$ for Fig.~\ref{fig:discoveryMCHM}.
We define a point in this space to be `discoverable' at a certain luminosity ${\cal L}$ if the probability of having a total number of events smaller than
or equal to $r_b {\cal L}$ is below $1\%$,
\begin{equation}\label{eq:discovery_condition}
\int_0^{r_b} \! d r'  \; p_{\cal L}(r' | (r_s+r_b) {\cal L}) \leq 0.01\, ,
\end{equation}
and if the number of observed events $(r_s + r_b) {\cal L}$ is 5 or larger. The discovery luminosity is thus defined to be the smallest value of ${\cal L}$ which satisfies 
these two conditions.

In the case of the plot of Fig.~\ref{fig:precision}, we marginalize the probability function over all possible values of the coupling $c$ and of 
the branching fraction  $BR(hh\to \gamma\gamma b\bar b)$ assuming that they  have a Gaussian distribution around their central value with 
20\% relative error. In practice, we set 
\begin{equation}
r_s(c_2,d_3,\theta_1,\theta_2) \equiv (1 + \delta_1 \theta_1)  \, BR(hh\to \gamma\gamma b\bar b) \, \sigma(\bar c\, (1 + \delta_2 \theta_2), c_2,d_3)\, ,
\end{equation}
where $\sigma(c,c_2,d_3)$ is the signal production cross section,
$\delta_{1,2}=0.20$ and $\bar c$ denotes the central value of $c$, and 
integrate over the nuisance parameters $\theta_{1,2}$:
\begin{equation}
p_{\cal L}(c_2, d_3 | N) \propto \int \!\! d\theta_1 \int \!\! d\theta_2 \  e^{-\theta_1^2/2} e^{-\theta_2^2/2}  \,
L(N | (r_s(c_2,d_3,\theta_1,\theta_2)+r_b)  {\cal L})\, \pi(c_2,d_3)
\, .
\end{equation}
The prior distribution is assumed to be flat over the  plane $(c_2,d_3)$
and the posterior probability  is normalized so that $\int\! dc_2 \!\int\! dd_3 \; p_{\cal L}(c_2, d_3 | N) =1$.

Finally, the plot of Fig.~\ref{fig:precisionMCHM} has been derived by expressing the posterior probability as a function of $\xi$
and marginalizing over all possible values of the branching fraction:
\begin{equation}
p_{\cal L}(\xi | N) \propto \int\!\! d\theta_1  \  e^{-\theta_1^2/2} \, L(N | ( (1 + \delta_1 \theta_1) \, r_s(\xi)+r_b)  {\cal L})\, \pi(\xi)\, .
\end{equation}
We choose a flat prior for $0 \leq \xi \leq 1$ (vanishing otherwise) and normalize the probability so that $\int_0^1 \! d \xi \; p_{\cal L}(\xi | N) =1$.
For each value $\xi_{th}$ the number of observed events has been set to the total 
expected value $N = (r_s(\xi_{th}) + r_b){\cal L}$.

\end{document}